\documentclass
[prl,reprint,showpacs,showkeys,final,nobibnotes,nofootinbib,onecolumn,11pt,superscriptaddress]{revtex4}%
\usepackage{amsmath}
\usepackage{amsfonts}
\usepackage{amssymb}
\usepackage{graphicx}
\usepackage{rotating}
\usepackage{appendix}%
\begin{document}
\title{A coefficient average approximation towards Gutzwiller
wavefunction formalism}
\author{Jun Liu}
\email{jun.physics@gmail.com}
\author{Yongxin Yao}
\author{Cai-Zhuang Wang}
\author{Kai-Ming Ho}
\affiliation{Ames Laboratory, US DOE and Department of Physics and Astronomy, Iowa State
University, Ames, Iowa 50011, USA}
\date{January, 26th, 2015}

\begin{abstract}
Gutzwiller wavefunction is a physically well motivated trial wavefunction for
describing correlated electron systems. In this work, a new approximation is
introduced to facilitate evaluation of the expectation value of any operator
within the Gutzwiller wavefunction formalism. The basic idea is to make use of
a specially designed average over Gutzwiller wavefunction coefficients
expanded in the many-body Fock space to approximate the ratio of expectation
values between a Gutzwiller wavefunction and its underlying noninteracting
wavefunction. To check with the standard Gutzwiller approximation (GA), we
test its performance on single band systems and find quite interesting
properties. On finite systems, we noticed that it gives superior performance
than GA, while on infinite systems it asymptotically approaches GA. Analytic
analysis together with numerical tests are provided to support this claimed
asymptotical behavior. At the end, possible improvements on the approximation
and its generalization towards multiband systems are illustrated and discussed.

\end{abstract}
\maketitle

\section{Introduction}

Correlation effects play an important role in electronic movements and
physical properties of real materials\cite{CORRELEC.2003}. Strong
electron-electron interaction is believed to be key to fully understand many
interesting phenomena, including high Tc superconductivity\cite{HighTC.1997},
heavy fermion behaviors\cite{HeavyFermion.2007}, abnormal transport and
optical properties\cite{OPTIC_RevModPhys.83.471}, and more on strongly
correlated materials. A versatile and convenient way to describe these systems
has been posed as a big challenge to the solid state community in the past
years. Part of the reasons are due to the strong Coulomb interaction
preventing a controlled perturbative treatment which has been very
successfully developed for weakly interacting systems. The conventional mean
field treatment is most versatile in gaining insights to correlated electron
systems on their possible new physics, competing phases and dynamical
behaviors. But its conclusions are always in question whether the ignored
residual electron correlation effects are still significant enough to
overshadow the presumed mean-field behaviors\cite{PhysRevLett.92.226402}%
\cite{PhysRevLett.94.127003}. To go beyond the mean field approximation,
analytic tools were proposed based on infinite summation of Feynmann diagrams
of chosen types, e.g., the Random Phase Approximation
(RPA)\cite{QUANTHEORY.2003}, the spin fluctuation theory\cite{SpinFluc.2008},
and compact diagrammatic equations like the Fluctuation Exchange approximation
(FLEX)\cite{AnnPhys.193.206} or the parquet
formalism\cite{SolidStateCommun.82.311}\cite{Parquet.2004}. Although they
offer alternative ways for people to use, their results might be biased by
starting with a subjective choice of a subgroup of diagrams which are usually
numerically convenient to deal with. On the side of computational physics, a
number of tools exist including Exact Diagonalization
(ED)\cite{ED_PhysRev.135.A640}, Quantum Monte Carlo
simulation(QMC)\cite{1st_QMC_PhysRevB.26.5033}, Dynamic Mean Field Theory
(DMFT) and its cluster extensions\cite{DMFT_RevModPhys.68.13}%
\cite{CDMFT_PhysRevLett.87.186401}\cite{DCA_RevModPhys.77.1027}, and
renormalization group type methods\cite{NRG_RevModPhys.47.773}%
\cite{NRG_RevModPhys.80.395}\cite{DMRG_PhysRevLett.69.2863}. These tools are
computationally demanding and can suffer from serious finite size effects or
other numerical complications.

On the other hand, the variational approach\cite{QUANMECH.2000} has always
been a very important category of methods in addressing a wide range of
physical problems, thanks to its capability of conveniently incorporating
clear physical insights into a well-designed trial wavefunction. A typical
example is the \textit{ab initio} local density approximation (LDA) which has
become an indispensable tool in modern scientific and material
research\cite{HK_PhysRev.136.B864}\cite{KS_PhysRev.140.A1133}. In strongly
correlated systems, the variational approach can play an important role and
has been shown to be very effective and enlightening in studying various
correlation effects\cite{RVB_PhysRevLett.58.2790}\cite{RVB_Science.235.1196}%
\cite{PhysRevLett.10.159}. It avoids looking for a small quantity to validate
a perturbative expansion of the problem, but instead, it focuses directly on
most prominent physics out of correlation effects. A simple-minded way to take
care of the local onsite correlation is to form a trial wavefunction with a
local projection operator to control double occupancy. This brings up the
famous Gutzwiller trial wavefunction to address local correlation
effects\cite{PhysRevLett.10.159}. Unfortunately, a direct evaluation of
operator expectation values is not practical for a many-body wavefunction. How
to form a new way to efficiently implement the Gutzwiller wavefunction
formalism will be the focus of the current paper.

The typical model Hamiltonian to study correlation effects is the so-called
single band Hubbard model\cite{ProcRoySocA.276.238} on a two-dimensional
square lattice, expressed as%
\begin{equation}
\hat{H}=-t\sum_{\left\langle i,j\right\rangle ,\sigma}c_{i,\sigma}^{\dagger
}c_{j,\sigma}+U\sum_{i}n_{i,\uparrow}n_{i,\downarrow}%
\end{equation}
with $i,j$ denoting site indices and $\sigma$ spins. Two energy scales, the
hopping amplitude, $t,$ between nearest neighboring sites and the onsite
Hubbard interaction, $U,$ exist and compete with each other. The correlation
effect becomes dominant in case onsite energy $U$ is big as compared against
hopping $t.$ The ground state wavefunction is presumably well described by the
Gutzwiller trial wavefunction (GWF) defined as
\begin{equation}
\left\vert \Psi\right\rangle _{G}=\left(  \prod_{i}\hat{P}_{i,G}\right)
\left\vert \Psi_{0}\right\rangle \label{1-0}%
\end{equation}
with the local Gutzwiller projector $\hat{P}_{i,G}$ defined as
\begin{equation}
\hat{P}_{G}=g^{\hat{n}_{\uparrow}\hat{n}_{\downarrow}} \label{1-0a}%
\end{equation}
Here $\hat{n}_{\uparrow}\hat{n}_{\downarrow}$ is defined as the local double
occupancy operator and $g$ is the unknown Gutzwiller variational parameter, a
positive number which controls the weight of electronic configurations
containing double occupancy in the noninteracting wavefunction $\left\vert
\Psi_{0}\right\rangle $\cite{PhysRev.134.A923}\cite{PhysRevLett.10.159}.
Although $\hat{P}_{G}$ is defined purely locally, it might still capture some
nonlocal physics through the 2nd order virtual hopping process in the strong
correlation limit\cite{HUBBARD_t-J_PhysRevB.37.533}\cite{HUBBARD_t-J.1990}.
Thus Eq. \ref{1-0} is a quite reasonable trial wavefunction to capture the
essential physics of a correlated electron system.

However, introducing a physically sensible trial wavefunction is only the
first step. Brute force evaluation of physical observables in a many-body
system can be a real pain due to numerical difficulties and computer hardware
capacity. Thus, the so-called Gutzwiller approximation (GA) was introduced to
facilitate the evaluation\cite{PhysRevLett.10.159}\cite{RevModPhys.56.99}%
\cite{Ogawa_GA_ProgTheorPhys.53.614}\cite{Bunemann_GA_EurPhysJB.4.29}. There,
the expectation value for an interacting system is assumed to be proportional
to that of the noninteracting wavefunction through a site-decomposeable
renormalization factor. Taking the hopping operator as an example, GA gives%
\begin{equation}
\left\langle c_{i,\alpha,\sigma}^{\dagger}c_{j,\beta,\sigma}\right\rangle
=z_{i,\alpha}z_{j,\beta}\left\langle c_{i,\alpha,\sigma}^{\dagger}%
c_{j,\beta,\sigma}\right\rangle _{0} \label{1-1}%
\end{equation}
for two distinct sites. Here, $\alpha,\beta$ denote orbital indices and
$\sigma$ denotes spin. A succinct analytic expression for $z_{i,\alpha}$ is
available for the operator $c_{i,\alpha,\sigma}^{\dagger}$ (or $c_{i,\alpha
,\sigma}$)\cite{Bunemann_GA_EurPhysJB.4.29}. Actually, we might come up with
an educated guess that, given any local operator $\hat{o}_{I}$ with $I$ the
set of local spin-orbital states involved in $\hat{o}$, the corresponding
$z_{\hat{o}}$ is given as%
\begin{equation}
z_{\tilde{o}}=\dfrac{1}{\sqrt{\prod_{\left(  \alpha,\sigma\right)  \in
I}n_{\left(  \alpha,\sigma\right)  }^{0}\left(  1-n_{\left(  \alpha
,\sigma\right)  }^{0}\right)  A_{\left(  \alpha,\sigma\right)  }}}\left(
\sum_{\Gamma,\Gamma^{\prime}}\sqrt{p_{\Gamma}p_{\Gamma^{\prime}}}\left\vert
\left\langle \Gamma^{\prime}\right\vert \hat{o}_{I}\left\vert \Gamma
\right\rangle \right\vert ^{2}\right)  \label{1-1a}%
\end{equation}
with
\begin{equation}
A_{\left(  \alpha,\sigma\right)  }=\left\{
\begin{array}
[c]{cc}%
\dfrac{n_{\left(  \alpha,\sigma\right)  }^{0}}{1-n_{\left(  \alpha
,\sigma\right)  }^{0}} & \text{if }\hat{n}_{\left(  \alpha,\sigma\right)
}\text{ is part of }\tilde{o}_{I}\\
1 & \text{o.w.}%
\end{array}
\right.
\end{equation}
Here $n_{\left(  \alpha,\sigma\right)  }^{0}=\left\langle \Psi_{0}\right\vert
\hat{n}_{\left(  \alpha,\sigma\right)  }\left\vert \Psi_{0}\right\rangle $ and
$p_{\Gamma}$ denotes the occupational probability at a local Fock state
$\Gamma.$ The summation in Eq. \ref{1-1a} traverses all local Fock states
mutually related by the local operator $\hat{o}_{I}.$ One can go through the
tedious yet well-established procedure to verify this relations for any
operator\cite{IDL-MBGA_JPhysCondensMatter.9.7343}. The introduction of
$z_{i,\alpha}$ as a renormalization factor is physically very sensible. Strong
correlation effects among electrons necessarily leave an impact on their
motions, thus modifying their dynamical and transport behaviors as compared
against a Fermi liquid system described by $\left\vert \Psi_{0}\right\rangle $.

There have been roughly three different ways to reach GA. Metzner et al
approached GA with mathematical rigor by applying Feymann diagrammatic
expansion techinque to the Gutzwiller wavefunction\cite{PhysRevLett.62.324}%
\cite{PhysRevB.37.7382}. The approach effectively cuts intersite
communications in the infinite spatial dimension limit and leaves only
intrasite correlations among local orbits. This brings up the main site-wise
decomposibility feature of GA readily seen from the site-dependent
renormalization factors illustrated in Eq. \ref{1-1}. This approach has been
further carried out by Bunemann et. al. and Fabrizio et. al. to multiband
systems and with a more generalized Gutzwiller
projector\cite{Bunemann_GA_EurPhysJB.4.29}\cite{MBGA_GONSITE_PhysRevB.57.6896}
\cite{JPhysSocJapan.65.1056}\cite{Fabrizio_MBGA_PhysRevB.76.165110}. The
second way of writing down GA is through physical intuition with a hand-waving
argument. Gutzwiller originally formulated GA based on some
assumptions\cite{PhysRevLett.10.159} whose physics were rather obscure thus
preventing it from being generalized to multiband systems. Ogawa's counting
argument formally pointed out that the physics underlying Gutzwiller's GA is
to assume that the expectation value of a product of number operators equals
the product of expectation value of each number operator in the
series\cite{Ogawa_GA_ProgTheorPhys.53.614}. The physics of GA is most
transparent in Bunemann's version of counting
argument\cite{Bunemann_GA_EurPhysJB.4.29}. We regard it as a third approach
towards GA due to its balance between clarity in physics and completeness in
formulation. Bunemann made it clear that the projection of a noninteracting
wavefunction onto a specific electronic configuration depends only on the
number of electrons on each local orbital on the lattice, but not on how these
electrons are distributed on the lattice. This implies two things. First, the
combinatoric trick accompanying the assumption can be used to evaluate any
inner product, which greatly simplifies the whole formalism and introduces an
additional convenience of taking the thermodynamic limit to further simplify
the expressions. Second, he pointed out that GA actually implies more than
just one relation, but instead, infinitely many for different operators whose
expectation values are to be evaluated. This leaves us a taste on how crude GA
might work in reality.

Despite many achievements made by GA on qualitatively addressing the Mott
physics in correlated electron systems, quantitatively however, it might
introduce artifacts. For instance, GA predicts existence of a Brinkman-Rice
metal-insulator transition at a critical onsite interaction comparable to the
bandwidth in a single band system\cite{BR-MIT_PhysRevB.2.4302}, while GWF
formalism should always give metallic behavior unless $U\rightarrow\infty
$\cite{PhysRevLett.62.324}. These artifacts are of course closely related with
the rough assumptions made in GA. The artificial Brinkman-Rice transition
seems to imply excessive local correlations were introduced by GA. In
addition, besides the unphysical infinite spatial dimension limit, the most
questionable assumption in GA is that it calls for a presumably known
connection between the Gutzwiller local orbital occupations and those of the
noninteracting wavefunction, a requisite to facilitate the Feynman
diagrammatic expansion to derive GA\cite{Bunemann_GA_EurPhysJB.4.29}%
\cite{IDL-MBGA_JPhysCondensMatter.9.7343}. This might cause trouble in
multiband systems where charge flow among orbitals might necessarily play a
role due to correlation effects\cite{OrbitSelecMT_EurPhyJB.25.191}.

It would be useful if some of the concerns mentioned above could be addressed.
However, within the existing GA formulation, this is not an easy task mainly
because the formulation is mathematically prohibitively complicated to be
further improved. Thus we need a new way to look into the renormalization
factors of operators, central quantity of GA to bridge between the Gutzwiller
and the noninteracting wavefunction. We come up with a very simple way to
design the renormalization factors which gives the standard GA in the
thermodynamic limit, and which also has the potential to be improved
systematically. We might thus call it a fourth way to reach GA.

\section{Method}

Let's introduce some notations first to facilitate the up-coming discussion.
Given a number of local spin-orbital states, the local Fock space is set up
and denoted as $\Gamma.$ Specifically, in the single band case, we denote the
spin-orbital states as $1=\left(  \alpha,\uparrow\right)  ,2=\left(
\alpha,\downarrow\right)  $ with $\alpha$ the orbital index, and note
$\Gamma\in\{\emptyset,(1),(2),\left(  1,2\right)  \}.$ Here $\emptyset$
denotes an empty state and $\left(  1\right)  $ denotes a singly occupied
state with spin-orbital state $1$ taken. Similar interpretation applies to
$\left(  2\right)  $ and $\left(  1,2\right)  $. Let's denote $\left\vert
\Gamma\right\vert $ to represent the number of electrons in a Fock state
$\Gamma,$ and $s_{z}\left(  \Gamma\right)  $ the total spin z component of
$\Gamma.$ Specifically, we have the following
\begin{align*}
\left\vert \emptyset\right\vert  &  =0,s_{z}\left(  \emptyset\right)  =0;\\
\left\vert \left(  1\right)  \right\vert  &  =1,s_{z}\left(  1\right)
=\frac{1}{2};\\
\left\vert \left(  2\right)  \right\vert  &  =1,s_{z}\left(  2\right)
=-\frac{1}{2};\\
\left\vert \left(  1,2\right)  \right\vert  &  =2,s_{z}\left(  1,2\right)  =0
\end{align*}
Suppose the system has a total of $N$ sites, $N_{e}$ electrons and $S_{z}$ net
spins. Each electronic configuration is described in an occupation
representation as $\left\{  \mathbf{\Gamma}\right\}  =\left\{  \Gamma
_{1},\Gamma_{2},\ldots,\Gamma_{N}\right\}  $ with subscripts denoting site
indices. These occupation representations form a complete Fock space on the
lattice, satisfying
\begin{equation}
\sum_{\left\{  \mathbf{\Gamma}\right\}  }\left\vert \mathbf{\Gamma
}\right\rangle \left\langle \mathbf{\Gamma}\right\vert =I \label{1-4}%
\end{equation}
The count of each local Fock state in $\mathbf{\Gamma,}$ $n_{\Gamma}\left(
\mathbf{\Gamma}\right)  ,$ is defined to be
\begin{equation}
n_{\Gamma}\left(  \mathbf{\Gamma}\right)  =\sum_{i}n_{i,\Gamma}\left(
\Gamma\right)  \text{ with }n_{i,\Gamma}\left(  \Gamma\right)  =\delta
_{\Gamma,\Gamma_{i}} \label{1-4a}%
\end{equation}
or simply $n_{\Gamma}$ if no confusion arouses in its interpretation. All
possible values for $n_{\Gamma}$ form a set denoted as $\left\{  n_{\Gamma
}\right\}  .$

Here is something quite general and useful for us to know. Consider a generic
trial wavefunction $\left\vert \Psi\right\rangle $ and note the closure
relation of Eq. \ref{1-4}, the inner product involving any operator $\hat{O}$
is written as%
\begin{equation}
\left\langle \Psi\right\vert \hat{O}\left\vert \Psi\right\rangle
=\sum_{\left\{  \mathbf{\Gamma}\right\}  ,\left\{  \mathbf{\Gamma}^{\prime
}\right\}  }\left\langle \Psi|\mathbf{\Gamma}\right\rangle \left\langle
\mathbf{\Gamma}\right\vert \hat{O}\left\vert \mathbf{\Gamma}^{\prime
}\right\rangle \left\langle \mathbf{\Gamma}^{\prime}|\Psi\right\rangle
\label{1-5}%
\end{equation}
This can be reexpressed in terms of a noninteracting wavefunction $\left\vert
\Psi_{0}\right\rangle $ on which $\left\vert \Psi\right\rangle $ is set up as
\begin{equation}
\left\langle \Psi\right\vert \hat{O}\left\vert \Psi\right\rangle
=\sum_{\left\{  \mathbf{\Gamma}\right\}  ,\left\{  \mathbf{\Gamma}^{\prime
}\right\}  }C\left(  \mathbf{\Gamma},\mathbf{\Gamma}^{\prime}\right)
\left\langle \Psi_{0}|\mathbf{\Gamma}\right\rangle \left\langle \mathbf{\Gamma
}\right\vert \hat{O}\left\vert \mathbf{\Gamma}^{\prime}\right\rangle
\left\langle \mathbf{\Gamma}^{\prime}|\Psi_{0}\right\rangle \label{1-6}%
\end{equation}
with
\begin{equation}
C\left(  \mathbf{\Gamma},\mathbf{\Gamma}^{\prime}\right)  =\frac{\left\langle
\Psi|\mathbf{\Gamma}\right\rangle }{\left\langle \Psi_{0}|\mathbf{\Gamma
}\right\rangle }\frac{\left\langle \mathbf{\Gamma}^{\prime}|\Psi\right\rangle
}{\left\langle \mathbf{\Gamma}^{\prime}|\Psi_{0}\right\rangle } \label{1-5a}%
\end{equation}
which holds as long as the denominators do not vanish, a fairly mild
constraint to be met in most cases. If we manage to replace $C\left(
\mathbf{\Gamma},\mathbf{\Gamma}^{\prime}\right)  $ with some constant $\bar
{C},$ an idea borrowed from the first mean value theorem of integrals, Eq.
\ref{1-6} is now
\begin{align}
\left\langle \Psi\right\vert \hat{O}\left\vert \Psi\right\rangle  &
\simeq\bar{C}\sum_{\left\{  \mathbf{\Gamma}\right\}  ,\left\{  \mathbf{\Gamma
}^{\prime}\right\}  }\left\langle \Psi_{0}|\mathbf{\Gamma}\right\rangle
\left\langle \mathbf{\Gamma}\right\vert \hat{O}\left\vert \mathbf{\Gamma
}^{\prime}\right\rangle \left\langle \mathbf{\Gamma}^{\prime}|\Psi
_{0}\right\rangle \nonumber\\
&  =\bar{C}\left\langle \Psi_{0}\right\vert \hat{O}\left\vert \Psi
_{0}\right\rangle \label{1-6-1}%
\end{align}
In the like fashion, the expectation value of $\hat{O}$ can be related to that
of the noninteracting system through
\begin{equation}
\left\langle \hat{O}\right\rangle =\frac{\left\langle \Psi\right\vert \hat
{O}\left\vert \Psi\right\rangle }{\left\langle \Psi|\Psi\right\rangle }%
\simeq\frac{\mathcal{F}}{\mathcal{B}}\frac{\left\langle \Psi_{0}\right\vert
\hat{O}\left\vert \Psi_{0}\right\rangle }{\left\langle \Psi_{0}|\Psi
_{0}\right\rangle }=Z_{\hat{O}}\left\langle \hat{O}\right\rangle _{0}
\label{1-7}%
\end{equation}
with
\begin{equation}
Z_{\hat{O}}=\frac{\mathcal{F}}{\mathcal{B}} \label{1-7a}%
\end{equation}
Here $\mathcal{F}$ and $\mathcal{B}$ are specifically chosen symbols to denote
the constant prefactors in the numerator and the denominator respectively.
Just like GA\cite{Bunemann_GA_EurPhysJB.4.29}, we call $Z_{\hat{O}}$ the
renormalization factor for operator $\hat{O}.$ Thus, through a renormalization
factor, the expectation value of an operator evaluated with a correlated
Gutzwiller wavefunction is directly related to that of a noninteracting
wavefunction. In principle, Eq. \ref{1-7} is able to rigorously hold if
$\mathcal{F}$ and $\mathcal{B}$ are chosen correctly. However, this is most
likely not the case in practice. How well Eq. \ref{1-7} holds depends on the
specific method used to set up $Z_{\hat{O}}.$

We believe the implication of the form of a renormalization factor in Eq.
\ref{1-7} reaches far beyond GA in that it gives us a clear route to determine
$Z_{\hat{O}}$ without resorting to the very involved algebras from Feynmann
diagrams and the unphysical assumptions used in GA. One way to determine
$Z_{\hat{O}},$ or equivalently $\mathcal{F}$ and $\mathcal{B}$ , is to go
through $C\left(  \mathbf{\Gamma},\mathbf{\Gamma}^{\prime}\right)  $ directly
by averaging these wavefunction coefficients in the many-body Fock space. You
will see that, in the single band case, even a simple definition towards
$\mathcal{F}$ and $\mathcal{B}$ through a wavefunction coefficient average
readily gives GA as its limiting behavior. Here are the details for defining
$\mathcal{F}$ and $\mathcal{B}$.

For the Gutzwiller trial wavefunction defined in Eq. \ref{1-0} and Eq.
\ref{1-0a} with only density operators entering the Gutzwiller projector, one
possible way to define $\bar{C}$ for any operator $\hat{O}$ is
\begin{equation}
\bar{C}_{\hat{O}}\left(  g\right)  =\frac{\sum_{\left\{  \mathbf{\Gamma
}\right\}  ,\left\{  \mathbf{\Gamma}^{\prime}\right\}  }\left[  \prod
_{i}g^{n_{i,\Gamma_{D}}\left(  \mathbf{\Gamma}\right)  +n_{i,\Gamma_{D}%
}\left(  \mathbf{\Gamma}^{\prime}\right)  }\right]  \mathcal{S}\left(  \hat
{O};\mathbf{\Gamma},\mathbf{\Gamma}^{\prime}\right)  \delta_{\mathbf{\Gamma}%
}^{\left(  3\right)  }\delta_{\mathbf{\Gamma}^{\prime}}^{\left(  3\right)  }%
}{\sum_{\left\{  \mathbf{\Gamma}\right\}  ,\left\{  \mathbf{\Gamma}^{\prime
}\right\}  }\mathcal{S}\left(  \hat{O};\mathbf{\Gamma},\mathbf{\Gamma}%
^{\prime}\right)  \delta_{\mathbf{\Gamma}}^{\left(  3\right)  }\delta
_{\mathbf{\Gamma}^{\prime}}^{\left(  3\right)  }}\label{1-8}%
\end{equation}
where $n_{i,\Gamma_{D}}\left(  \mathbf{\Gamma}\right)  =\delta_{\Gamma
_{D},\Gamma_{i}}$ with $\Gamma_{D}=c_{\uparrow}^{\dag}c_{\downarrow}^{\dag
}\left\vert 0\right\rangle $ denoting the doubly occupied Fock state. The
delta functions,
\begin{equation}
\delta_{\mathbf{\Gamma}}^{\left(  3\right)  }=\delta\left(  \sum_{\Gamma
}n_{\Gamma}-N\right)  \delta\left(  \sum_{\Gamma}n_{\Gamma}\left\vert
\Gamma\right\vert -N_{e}\right)  \delta\left(  \sum_{\Gamma}n_{\Gamma}%
s_{z}\left(  \Gamma\right)  -S_{Z}\right)
\end{equation}
ensure conservation of total charge and spin, and expansion on all sites. The
indicator function $\mathcal{S}\left(  \hat{O};\mathbf{\Gamma},\mathbf{\Gamma
}^{\prime}\right)  $ is defined as
\begin{equation}
\mathcal{S}\left(  \hat{O};\mathbf{\Gamma},\mathbf{\Gamma}^{\prime}\right)
=\left\{
\begin{array}
[c]{cc}%
1 & \text{if }\left\langle \mathbf{\Gamma}\right\vert \hat{O}\left\vert
\mathbf{\Gamma}^{\prime}\right\rangle \neq0\\
0 & \text{o.w.}%
\end{array}
\right.
\end{equation}
Thus the operator $\hat{O}$ has its effect included through $\mathcal{S}%
\left(  \hat{O};\mathbf{\Gamma},\mathbf{\Gamma}^{\prime}\right)  .$ Although
looked complicated, Eq. \ref{1-8} is just a carefully designed average over
nonvanishing elements of the inner product $_{G}\left\langle \Psi\right\vert
O\left\vert \Psi\right\rangle _{G}$ expanded in the Fock space set up on the
whole lattice. Thus it readily guarantees the correct limiting behavior as $g$
approaches unity. Why $\bar{C}_{\hat{O}}$ can be conveniently written out in
Eq. \ref{1-8} is closely related to the current choice of the Gutzwiller
projector defined on number operators only. An additional advantage with such
a Gutzwiller projector is, this enables us to use combinatorics to simplify
the whole expression.

Now we are ready to define $\left\langle \hat{O}\right\rangle $ and its
renormalization factor $Z_{\hat{O}}.$ Without loss of generality, let's assume
$\hat{O}=\hat{O}_{ijkl}$ acts on a sublattice $\mathcal{R=}\left\{
i,j,k,l\right\}  $ defined on the affected sites, and denote the number of
distinct sites in $\mathcal{R}$ to be $N_{\mathcal{R}}.$ We define
$\mathcal{\tilde{R}}$ as the complementary lattice to $\mathcal{R}$ and denote
its number of sites to be $N_{\mathcal{\tilde{R}}}.$ A Fock space on the whole
lattice can be decomposed into Fock spaces on the sublattices $\mathcal{R}$
and $\mathcal{\tilde{R}}.$ The counts on local Fock states in both sublattices
are denoted as $n_{\Gamma}$ and $\tilde{n}_{\Gamma}$ respectively. Then, apply
Eq. \ref{1-8} to both numerator and denominator and we have%
\begin{equation}
\left\langle \hat{O}_{ijkl}\right\rangle \simeq\frac{\mathcal{F}}{\mathcal{B}%
}\left\langle \hat{O}_{ijkl}\right\rangle _{0}\label{1-9}%
\end{equation}
with
\begin{align}
\mathcal{F} &  \mathcal{=}\bar{C}_{\hat{O}}\left(  g\right)  =\frac
{\sum_{\left\{  \mathbf{\Gamma}\right\}  ,\left\{  \mathbf{\Gamma}^{\prime
}\right\}  }^{\prime}\left\{  \left(  \prod_{i\in\mathcal{R}}g^{n_{i,\Gamma
_{D}}\left(  \mathbf{\Gamma}\right)  +n_{i,\Gamma_{D}}\left(  \mathbf{\Gamma
}^{\prime}\right)  }\right)  \mathcal{S}\left(  \hat{O};\mathbf{\Gamma
},\mathbf{\Gamma}^{\prime}\right)  \left[  \sum_{\left\{  \tilde{n}_{\Gamma
}\right\}  }^{"}\mathcal{C}_{N_{\mathcal{\tilde{R}}}}^{\left\{  \tilde
{n}_{\Gamma}\right\}  }g^{2\tilde{n}_{\Gamma_{D}}}\tilde{\delta}^{\left(
3\right)  }\right]  \right\}  }{\sum_{\left\{  \mathbf{\Gamma}\right\}
,\left\{  \mathbf{\Gamma}^{\prime}\right\}  }^{\prime}\left\{  \mathcal{S}%
\left(  \hat{O};\mathbf{\Gamma},\mathbf{\Gamma}^{\prime}\right)  \left[
\sum_{\left\{  \tilde{n}_{\Gamma}\right\}  }^{"}\mathcal{C}%
_{N_{\mathcal{\tilde{R}}}}^{\left\{  \tilde{n}_{\Gamma}\right\}  }%
\tilde{\delta}^{\left(  3\right)  }\right]  \right\}  }\label{1-9a}\\
\mathcal{B} &  \mathcal{=}\bar{C}_{\hat{I}}\left(  g\right)  =\frac
{\sum_{\left\{  n_{\Gamma}\right\}  }\left\{  \mathcal{C}_{N}^{\left\{
n_{\Gamma}\right\}  }g^{2n_{\Gamma_{D}}}\delta_{\mathbf{\Gamma}}^{\left(
3\right)  }\right\}  }{\sum_{\left\{  n_{\Gamma}\right\}  }\left\{
\mathcal{C}_{N}^{\left\{  n_{\Gamma}\right\}  }\delta_{\mathbf{\Gamma}%
}^{\left(  3\right)  }\right\}  }\label{1-9b}%
\end{align}
Here $\hat{I}$ denotes the identity operator. The primed sum in Eq. \ref{1-9a}
denotes that $\left\{  \mathbf{\Gamma}\right\}  $ and $\left\{  \mathbf{\Gamma
}^{\prime}\right\}  $ are confined within sublattice $\mathcal{R}$ only. The
double primed sum assumes $\tilde{n}_{\Gamma}$ and $\tilde{n}_{\Gamma_{D}}$
are counted within sublattice $\tilde{R}$ for fixed $\mathbf{\Gamma}$. The
delta functions $\tilde{\delta}^{\left(  3\right)  }$ states the conservation
constraints reenforced on $\mathcal{\tilde{R}}$ are%
\begin{equation}
\tilde{\delta}^{\left(  3\right)  }=\delta\left(  \sum_{\Gamma}^{"}\tilde
{n}_{\Gamma}-N_{\tilde{N}}\right)  \delta\left(  \sum_{\Gamma}^{"}\tilde
{n}_{\Gamma}\left\vert \Gamma\right\vert -\tilde{N}_{e}\right)  \delta\left(
\sum_{\Gamma}^{"}\tilde{n}_{\Gamma}s_{z}\left(  \Gamma\right)  -\tilde{S}%
_{Z}\right)  \label{1-9c}%
\end{equation}
with
\begin{align}
\tilde{N}_{e} &  =N_{e}-N_{e}^{\mathcal{R}}\left(  \mathbf{\Gamma}\right)
\label{1-9d}\\
\tilde{S}_{Z} &  =S_{Z}-S_{Z}^{\mathcal{R}}\left(  \mathbf{\Gamma}\right)
\label{1-9e}%
\end{align}
where $N_{e}^{\mathcal{R}}\left(  \mathbf{\Gamma}\right)  $ and $S_{Z}%
^{\mathcal{R}}\left(  \mathbf{\Gamma}\right)  $ denote total number of
electrons and net spins on sublattice $\mathcal{R}$ occupied by a specific
electronic configuration $\mathbf{\Gamma}.$ Note, a physical operator would
not change electron occupation and total spin z component, which implies
$N_{e}^{\mathcal{R}}\left(  \mathbf{\Gamma}\right)  =N_{e}^{\mathcal{R}%
}\left(  \mathbf{\Gamma}^{\prime}\right)  $ and $S_{Z}^{\mathcal{R}}\left(
\mathbf{\Gamma}\right)  =S_{Z}^{\mathcal{R}}\left(  \mathbf{\Gamma}^{\prime
}\right)  $. $\mathcal{C}_{N}^{\left\{  n_{\Gamma}\right\}  }$ takes the
standard definition of the multinomial coefficient,%
\begin{equation}
\mathcal{C}_{N}^{\left\{  n_{\Gamma}\right\}  }=\frac{N!}{\prod_{\Gamma
}\left(  n_{\Gamma}!\right)  }%
\end{equation}
Eq. \ref{1-9} to Eq. \ref{1-9e} completes the expressions for approximating
the renormalization factors through wavefunction coefficient averaging. These
expressions can be directly used to approximate the energy expectation value
of a Hamiltonian of a small system.

For big systems, one needs to take the thermodynamic limit to simplify the
coefficients in Eq. \ref{1-9}. A typical sum involved in Eq. \ref{1-9a} and
Eq. \ref{1-9b} is%
\begin{equation}
A\left(  \left\{  \tilde{n}_{\Gamma}\right\}  \right)  =\sum_{\left\{
\tilde{n}_{\Gamma}\right\}  }^{"}\mathcal{C}_{N_{\mathcal{\tilde{R}}}%
}^{\left\{  \tilde{n}_{\Gamma}\right\}  }g^{2\tilde{n}_{\Gamma_{D}}}%
\delta^{\left(  3\right)  }\label{1-10}%
\end{equation}
In the thermodynamic limit, the sum can be replaced by a single term located
at some unknown $\left\{  \tilde{n}_{\Gamma}^{\ast}\right\}  $ where the terms
are peaked. To solve for $\left\{  \tilde{n}_{\Gamma}^{\ast}\right\}  $ we
introduce Lagrange multipliers to relax the constraints entering Eq.
\ref{1-10} and replace the factorials with their asymptotic expressions using
the Sterling's formula. We finally come up with the following functional%
\begin{align}
h\left(  \left\{  \tilde{n}_{\Gamma}\right\}  ,\alpha,\beta,\gamma\right)   &
=\tilde{N}\ln\tilde{N}-\sum_{\Gamma}\tilde{n}_{\Gamma}\ln\tilde{n}_{\Gamma
}+\sum_{\Gamma}2\tilde{n}_{\Gamma}\ln g_{\Gamma}\nonumber\\
&  +\alpha\left(  \sum_{\Gamma}^{\prime}\tilde{n}_{\Gamma}-\tilde{N}\right)
+\beta\left(  \sum_{\Gamma}^{\prime}\tilde{n}_{\Gamma}\left\vert
\Gamma\right\vert -\tilde{N}_{e}\right)  +\gamma\left(  \sum_{\Gamma}^{\prime
}\tilde{n}_{\Gamma}s_{z}\left(  \Gamma\right)  -\tilde{S}_{z}\right)
\end{align}
whose minimization gives $\left\{  \tilde{n}_{\Gamma}^{\ast}\right\}  .$ The
Gutzwiller variational parameter, $g,$ now carries a local Fock state index
for the purpose of formalism consistency. Following the standard procedure to
extremize $h\left(  \left\{  \tilde{n}_{\Gamma}\right\}  ,\alpha,\beta
,\gamma\right)  $ by taking partial derivatives with respect to $\left\{
\tilde{n}_{\Gamma}\right\}  $, we get explicit expressions for $\tilde
{n}_{\Gamma}^{\ast}$ as
\begin{equation}
\frac{\tilde{n}_{\Gamma}^{\ast}}{\tilde{N}}=\frac{g_{\Gamma}^{2}\exp\left(
\beta\left\vert \Gamma\right\vert +\gamma s_{z}\left(  \Gamma\right)  \right)
}{\sum_{\Gamma}g_{\Gamma}^{2}\exp\left(  \beta\left\vert \Gamma\right\vert
+\gamma s_{z}\left(  \Gamma\right)  \right)  }\label{1-11}%
\end{equation}
The remaining unknown, $\beta$ and $\gamma,$ satisfy the following two
nonlinear equations,%
\begin{align}
\sum_{\Gamma}\left(  \left\vert \Gamma\right\vert -\tilde{N}_{e}/\tilde
{N}\right)  g_{\Gamma}^{2}\exp\left(  \beta\left\vert \Gamma\right\vert
+\gamma s_{z}\left(  \Gamma\right)  \right)   &  =0\label{1-11a}\\
\sum_{\Gamma}\left(  s_{z}\left(  \Gamma\right)  -\tilde{S}_{z}/\tilde
{N}\right)  g_{\Gamma}^{2}\exp\left(  \beta\left\vert \Gamma\right\vert
+\gamma s_{z}\left(  \Gamma\right)  \right)   &  =0\label{1-11b}%
\end{align}
Note $\tilde{N}_{e}$ and $\tilde{S}_{z}$ defined in Eq. \ref{1-9d} and Eq.
\ref{1-9e} explicitly enter the above two equations. This implies that each
term of the primed sums in $\mathcal{F}$ and $\mathcal{B}$ will have its own
optimal $\left\{  \tilde{n}_{\Gamma}^{\ast}\right\}  $. This complicates the
whole computation, but is expected to have a mild impact on its performance as
Eq. \ref{1-11a} and Eq. \ref{1-11b} are usually well-behaved.

Eq. \ref{1-9} and its accompanying definition of Eq. \ref{1-9a} and Eq.
\ref{1-9b} looks fundamentally different from GA in both the underlying
assumptions and their mathematical formulations. Surprisingly, however, the
current scheme can be shown to be well related to GA in the single band
system. A brief proof is provided in Appendix A and a numerical study is given
in Fig. \ref{approach} in the main text. It clearly reveals that GA can be
regarded as being an overly simplified approximation of the current scheme,
but with vanishing difference at infinite lattice size. Thus, we have provided
here a fourth perspective to look into GA and its local nature accompanying
the infinite spatial dimension limit towards the GWF formalism. Furthermore,
the proof readily suggests that such a simple scheme has superior performance
than GA in finite systems. This fact is case studied here on small Hydrogen
clusters. To distinguish it from GA, we call it AA in the coming discussions.

\section{Comparison between the new approximation and GA}%

\begin{figure}
[ptb]
\begin{center}
\includegraphics[
height=2.8634in,
width=4.9727in
]%
{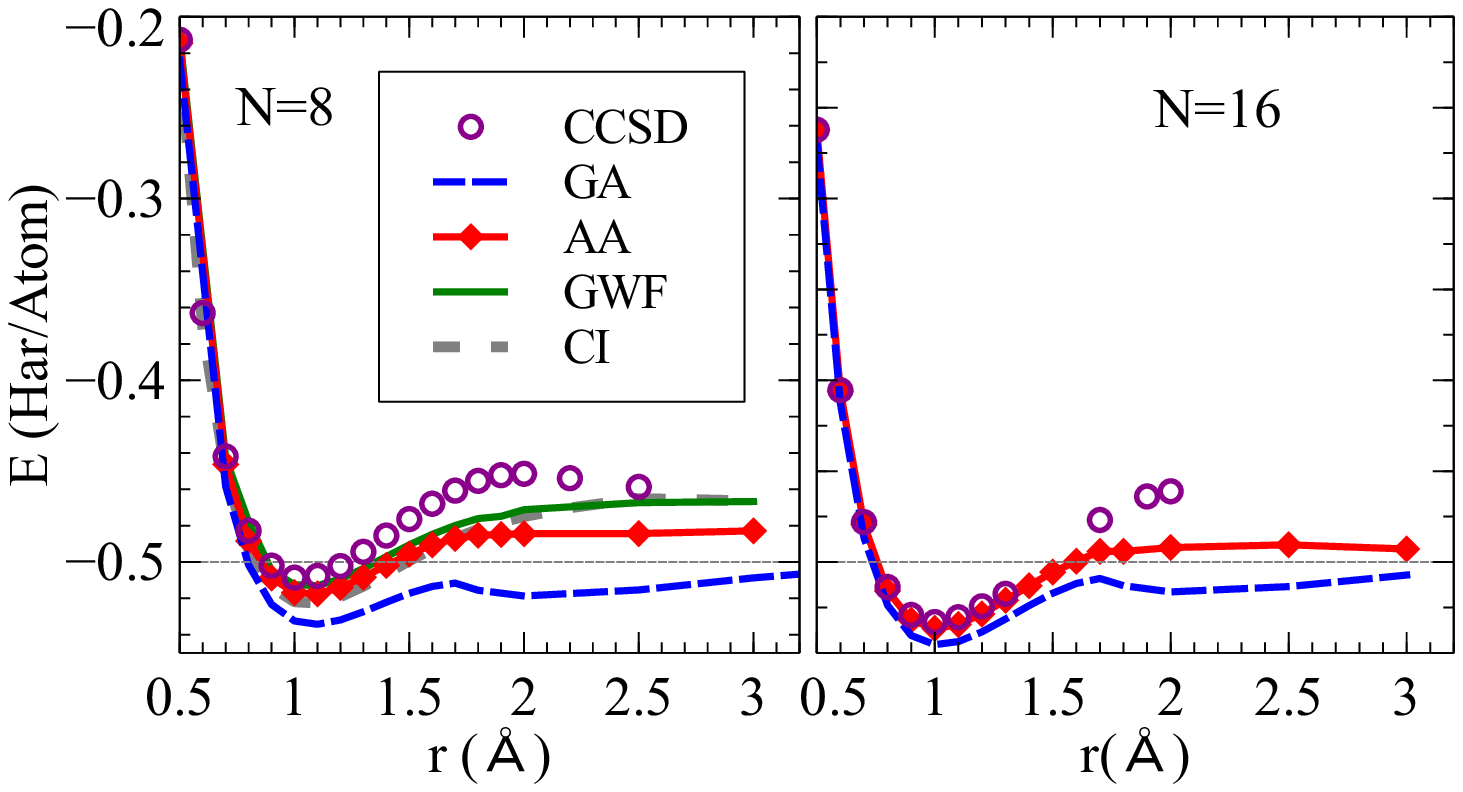}%
\caption{Total energy evaluated using the current method (denoted as AA), GA,
Coupled Cluster (CCSD), Configuration Interaction (CI), and GWF if possible,
on circular Hydrogen (H) chains of N=8 and N=16 atoms. The \textit{ab initio}
Hamiltonians are returned by GAMESS(US) using a STO-3G basis set description
of H. For N=8, GWF and CI results are both available to benchmark AA and GA.
For N=16, only CCSD is evaluated to check both approximations. Note CCSD does
not converge well beyond the bond breaking region. GWF and CI do not reach the
known Hydrogen atomic energy of -0.5Hartree due to the use of minimum basis
set. The noninteracting wavefunction underlying GWF (thus GA and AA as well)
are fixed at the Hartree-Fock solution.}%
\label{energy}%
\end{center}
\end{figure}
\begin{figure}
[ptbptb]
\begin{center}
\includegraphics[
height=3.3935in,
width=3.0113in
]%
{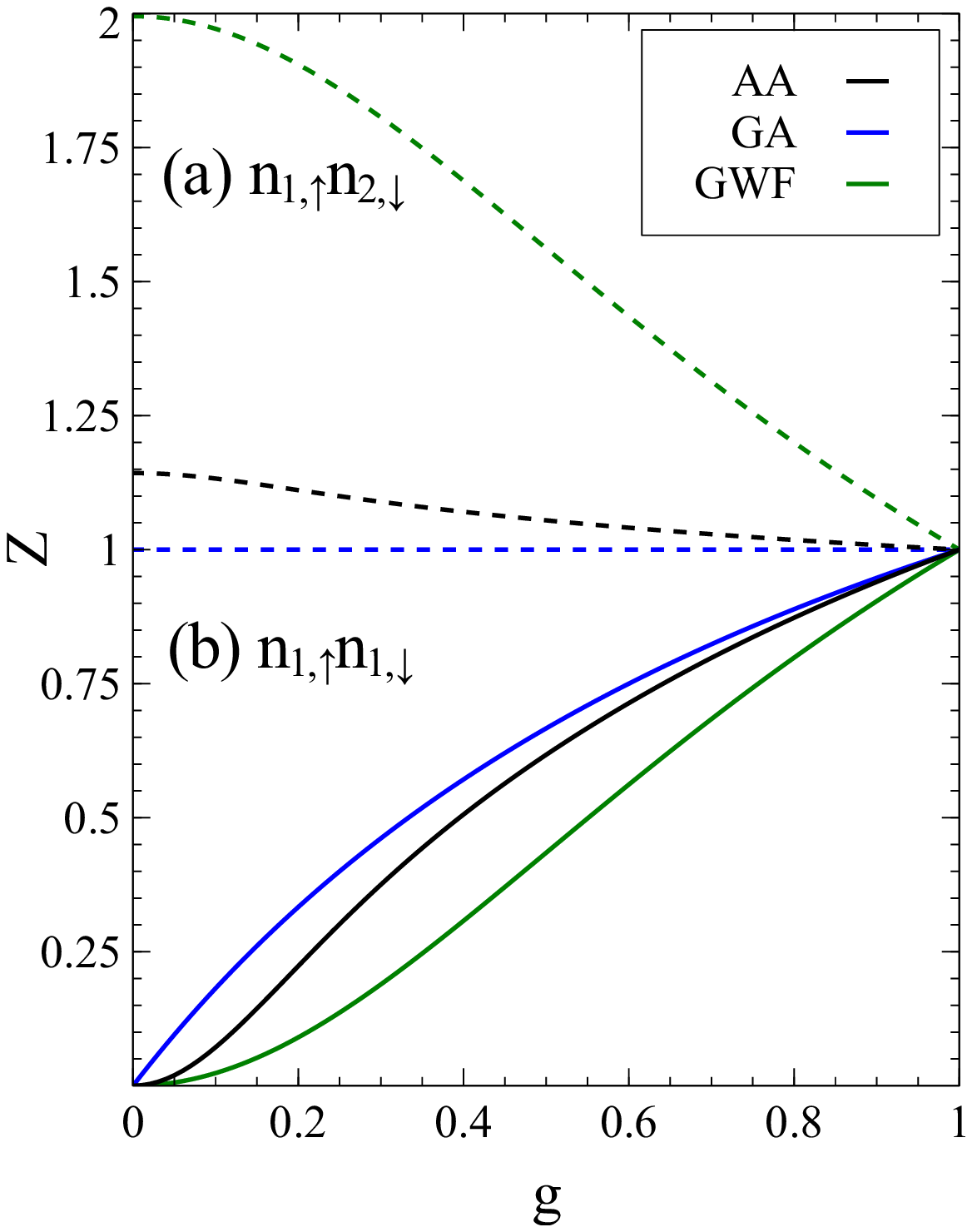}%
\caption{Shown in the figure are the renormalization factors $Z,$ defined in
Eq. \ref{1-7a}, as a function of $g$ for the onsite and nearest neighbor
density-density interactions evaluated with GWF, AA and GA. The calculation is
on a circular chain of $8$ Hydrogen atoms described with a STO-3G basis set
and separated with an atomic distance of $2.5\mathring{A}.$ The noninteracting
wavefunction is provided by a restricted Hartree Fock calculation. The optimal
Gutzwiller parameter has a value of $g\simeq0.02$. Energy from AA is about
$0.02$ $Har$ lower than GWF at $r=2.5\mathring{A}$ as seen from Fig
\ref{energy}. This energy difference is mainly due to the inaccurate
$\left\langle n_{1,\uparrow}n_{2,\downarrow}\right\rangle $ evaluated with AA,
the current approximation.}%
\label{zfac_nn}%
\end{center}
\end{figure}

Let's first look at the performance of AA and GA on predicting the ground
state energy of a finite system constructed by Hydrogen atoms. We choose the
one-dimensional minimum basis Hydrogen chains with periodic boundary
conditions to carry out the calculation. The \textit{ab initio} Hamiltonian
is
\begin{equation}
H=\sum_{\left(  i,j\right)  ,\sigma}t_{ij}c_{i,\sigma}^{\dagger}c_{j,\sigma
}+\sum_{\left(  i,j,k,l\right)  ;\sigma,\sigma^{\prime}}U_{i,j,k,l}%
c_{i,\sigma}^{\dagger}c_{j,\sigma^{\prime}}^{\dagger}c_{k,\sigma^{\prime}%
}c_{l,\sigma} \label{1-17}%
\end{equation}
where $\left(  i,j\right)  $ runs through all possible site pairs and $\left(
i,j,k,l\right)  $ describes all possible 2-body interactions on the chain. The
bare energy parameters $t$ and $U$ are evaluated with GAMESS (US), a Quantum
Chemistry package widely used for molecular calculations. Among all the
interactions, the density-density interactions are the most dominant. Its
ratio against nearest neighbor hopping is a quantitative measure of the
strength of correlation, which increases as inter-atomic distance increases.
The Gutzwiller wavefunction is constructed by applying the Gutzwiller
projector defined in Eq. \ref{1-0a} onto the noninteracting Hartree-Fock
wavefunction formed by the occupied molecular orbitals. We check the
performance of both approximations on two systems formed by $8$ and $16$
hydrogen atoms respectively. The brute force evaluation with the Gutzwiller
trial wavefunction (GWF) and the Configuration Interaction (CI) calculation
give benchmark results for AA and GA to compare. As shown in Fig.
\ref{energy}, GWF gives quite close energy to CI and is able to reach the
correct atomic energy given by CI. One might note that CI gives an atomic
Hydrogen ground state energy differring from the well-known value of
$-13.6eV.$ This is due to the STO-3G minimum basis set chosen to set up the
\textit{ab initio} Hamiltonian in Eq. \ref{1-17}. In the $16$ Hydrogen atom
system, GWF and CI are not convenient to evaluate. Thus, we use the Coupled
Cluster method(CCSD) instead to give an estimate on the ground state energy.
CCSD is a very popular method in Quantum Chemistry\cite{CCSD.2009} with good
balance between speed and accuracy. It has some known issues but is good
enough for our current purpose. The comparison between AA and GA, together
with other methods, are presented in Fig. \ref{energy}, where the ground state
energies are plotted against different inter-atomic separations. Several
things are quite interesting to note. GA is in general not as good as AA, but
its performance seems to be improved as system size increases. This gives a
strong support on our proof of GA being a grand canonical ensemble description
of the current coefficient averaging approximation, while AA is its canonical
ensemble description. Another thing to notice is that, as compared with GWF,
AA is slightly higher in energy around the bonding region and does not reach
the atomic limit energy returned by GWF. This is mainly due to biased onsite
double occupancy as well as insufficient enhancement of off-site
density-density correlations in the new approximation, as manifested in Fig
\ref{zfac_nn} for an inter-atomic separation of $2.5\mathring{A}$ on the $8$
Hydrogen chain system. Consider that the optimal Gutzwiller parameter $g$ is
quite small, AA thus gives roughly consistent onsite energy as GWF, but
underestimates the inter-site Coulomb repulsion. Consequently, AA gives an
underestimated bounding energy. On the other hand, GA deviates more seriously
from the benchmark energies. It also gives rise to an unphysical kink close to
the bond breaking regime hinting a plausible meta-stable state for the
Hydrogen rings.%

\begin{figure}
[ptb]
\begin{center}
\includegraphics[
height=3.7308in,
width=5.3255in
]%
{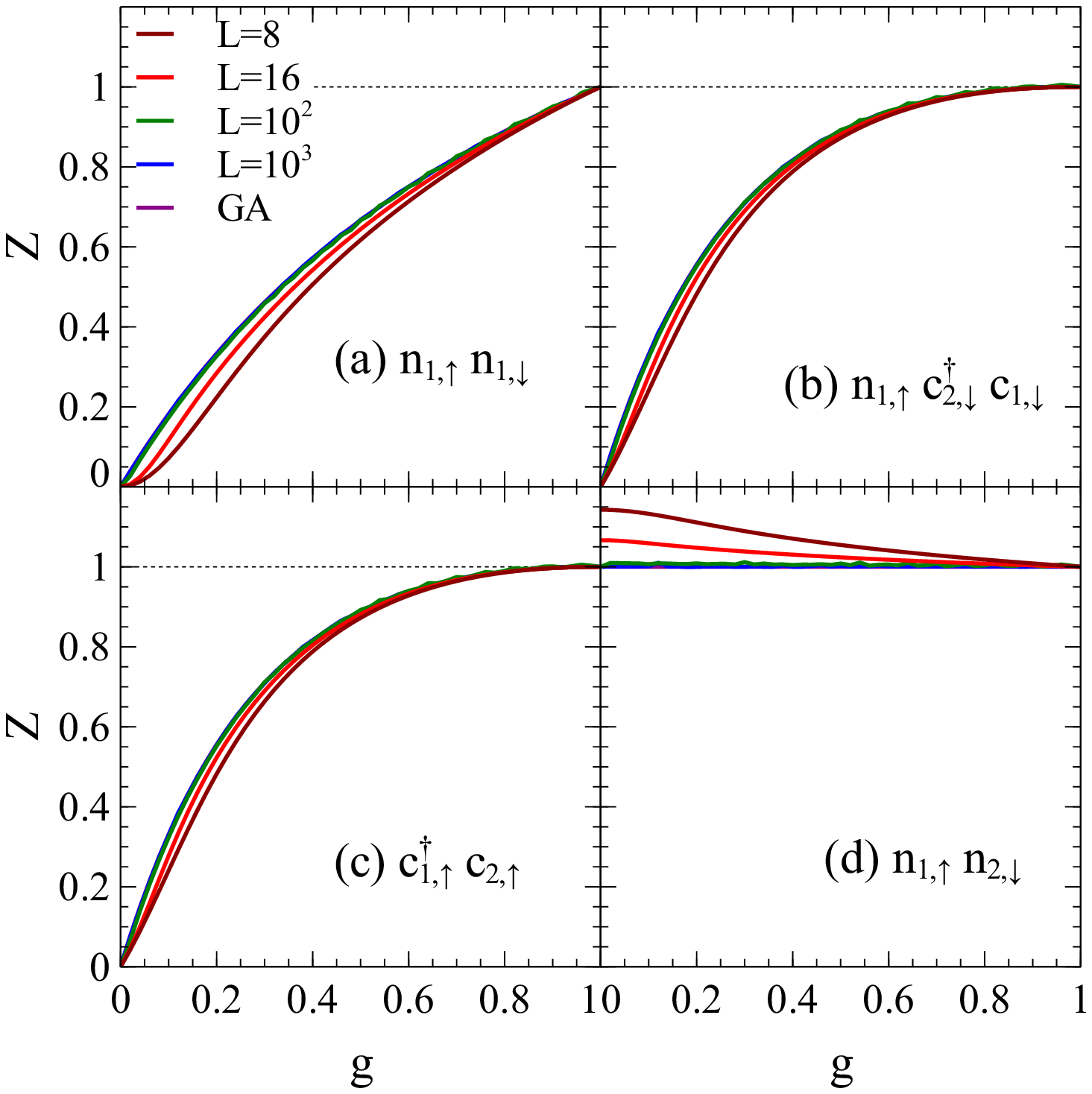}%
\caption{Renormalization factor $Z$ of four typical operators, $n_{1,\uparrow
}n_{1,\downarrow},$ $n_{1,\uparrow}c_{2,\downarrow}^{\dagger}c_{1,\downarrow
},$ $c_{1,\uparrow}^{\dagger}c_{2,\uparrow}$ and $n_{1,\sigma}n_{2,\sigma
^{\prime}}$ are evaluated with AA for different linear H chain sizes. GA
results are also provided for comparison. All graphs share the same legend
given in graph (a). Note, no Hamiltonian is needed to scan $g$ dependence of
AA and GA, only the Gutzwiller wavefunction is relevant.}%
\label{zfactor}%
\end{center}
\end{figure}

\begin{figure}
[ptb]
\begin{center}
\includegraphics[
height=2.8219in,
width=4.0283in
]%
{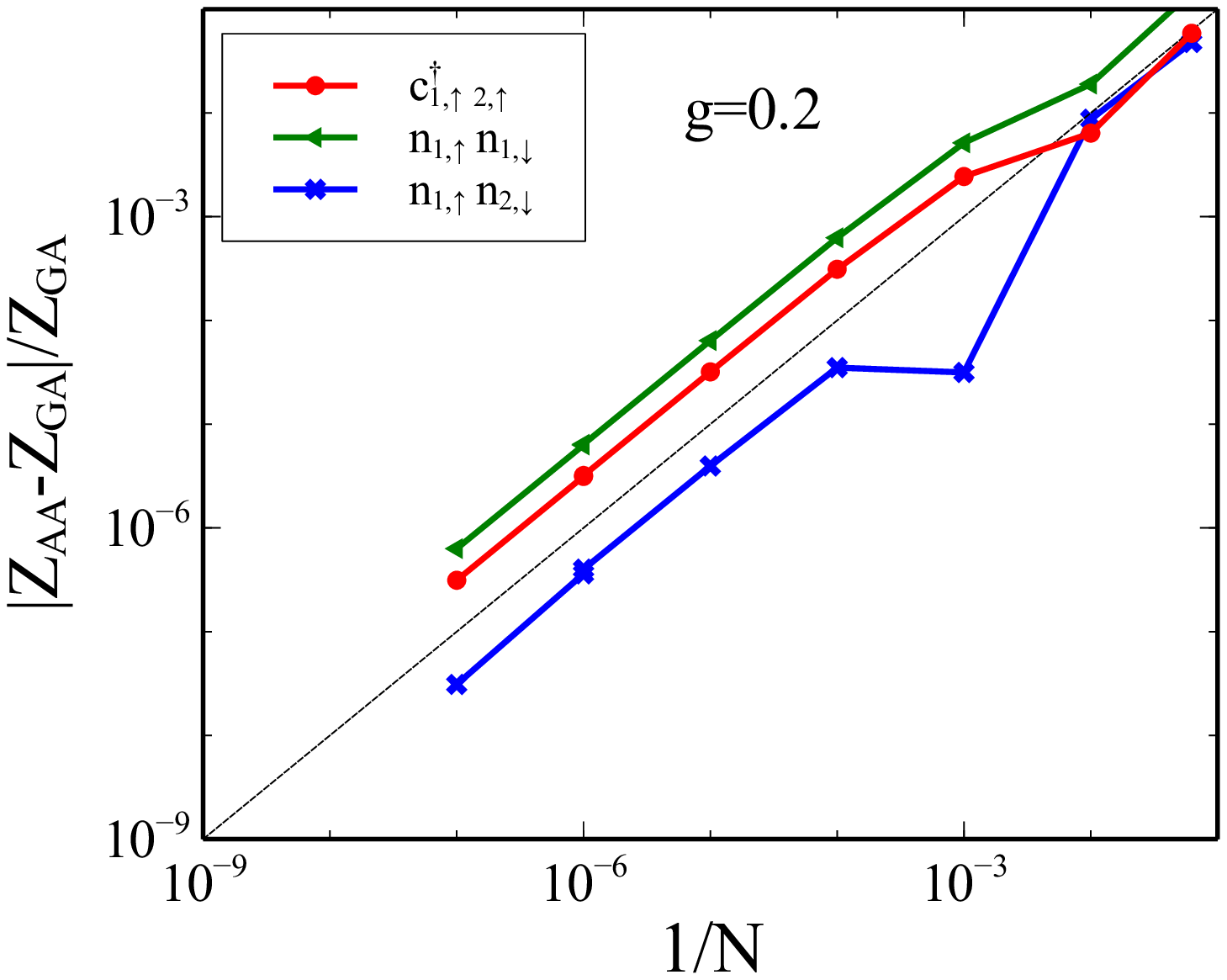}%
\caption{System size dependence of the relative renormalization factor
differences between the new approximation and GA is shown for 3 different
operators discussed in Fig. \ref{zfactor}. The calculation is carried out with
$g=0.2.$ $Z_{n_{1,\uparrow}c_{1,\downarrow}^{\dagger}c_{2,\downarrow}%
}=Z_{c_{1,\uparrow}^{\dagger}c_{2,\uparrow}}$ for both approximations and is
thus suppressed from the plot. Note the log scale is used on both axes.}%
\label{approach}%
\end{center}
\end{figure}

As the system size increases, renormalization factors, $Z_{\hat{O}},$ in the
new approximation gradually approach the GA results, as illustrated in Fig.
\ref{zfactor} and Fig. \ref{approach} on four typical operators in the
half-filled case. Results deviating from half-filling are not shown as both
methods are quite close to GWF results(e.g. see Ref
\cite{GA_GWF_Fill_PhysRevB.37.7382} for the nonhalf-filling behavior of GA).
From Fig. \ref{zfactor} we immediately see noticeable finite size effects for
AA. It converges quickly as system size increases. At a system size of
$N=10^{2},$ both AA and GA are nearly indistinguishable from each other. To
provide a more quantitative measure on the asymptotic behavior of the new
approximation, we present the system size dependence of the relative
differences of renormalization factors between AA and GA in Fig.
\ref{approach}. The manifest linear dependence between these two quantities
provides solid support on the conclusion that the new approximation
asymptotically approaches GA in its infinite lattice limit. This might also
suggest that the single band Gutzwiller approximation is quite an unexpectedly
stable limit for the Gutzwiller wavefunction formalism with a site-wise local
Gutzwiller projector.

\section{An attempt to apply the new approximation to multiband systems}

Thus far, we have applied the new approximation to single band systems and
compared it against GA and other benchmark calculations. Encouraged by the
limited success, we would like to generalize it to multiband systems which are
physically more interesting and relevant to real problems. The simplest way to
do so is to rewrite the Gutzwiller projector in terms of multi-orbital
particle number operators\cite{IDL-MBGA_JPhysCondensMatter.9.7343}, and to
follow Eq. \ref{1-9} to Eq. \ref{1-9e} to impose overall physical constraints.
Unfortunately, this is not successful due to the very fact that the proposed
approximation itself does not automatically preserve the correct total number
of electrons in the system except for the single band case. Indeed, there is
no guarantee that relevant physical constraints can be automatically preserved
with such a simple and arbitrary wavefunction coefficient averaging scheme. It
is a bit of luck that it does work out for single band systems.

However, this is not the excuse for us to give up the whole idea as it does
provide us with a candidate recipe to define the renormalization factor for
any operator. If we could look into the issue of failure a bit closer and at
the same time fill us with confidence that the idea of averaging does work
well on single band systems, we might come up with the following
interpretation on the validity of the scheme: the averaging procedure does
capture part of the correlation physics by giving reduced hopping or onsite
double occupancy, etc., but it might describe correlation physics only
qualitatively right in generic cases. One way to improve its quality requires
more degrees of freedom to enable adjusting the way how the averaging can be
carried out. This is actually quite similar to the logic Bunemann et. al.
followed in deriving the multiband Gutzwiller
approximation\cite{IDL-MBGA_JPhysCondensMatter.9.7343}%
\cite{Bunemann_GA_EurPhysJB.4.29}. There, a set of orbital-wise fugacity
parameters are introduced into the Gutzwiller trial wavefunction as adjustable
parameters to ensure that the orbital occupations take the values evaluated
with the noninteracting wavefunction\cite{IDL-MBGA_JPhysCondensMatter.9.7343},
or fulfill some constraints of similar nature in case the local Gutzwiller
projector is more general than involving only particle
operators\cite{MBGA_GONSITE_PhysRevB.57.6896}%
\cite{Fabrizio_MBGA_PhysRevB.76.165110}. Their purpose to choose these
constraints is for mathematical convenience to enable a rigorous Feynmann
diagrammatic expansion to the total energy expectation value evaluated with
the Gutzwiller trial wavefunction. Similarly, we will introduce the fugacity
parameters into the trial wavefunction and impose other more reasonable
physical constraints than creating a connection with information from the
noninteracting wavefunction. We will use these fugacity parameters to
calculate the renormalization factors defined through averaging. From a
statistical point of view, these fugacity parameters define a weight on each
wavefunction coefficient. Thus the renormalization factors are defined not as
a simple averaging, but as a weighted averaging instead. With a careful choice
of weights, a concept readily borrowed from the statistics theory, more
reasonable renormalization factors are anticipated to make this new
approximation useful. Specifically, we introduce into the trial wavefunction
of Eq. \ref{1-0} a local weighting operator
\begin{equation}
\hat{W}_{i}=\prod_{s}\left(  \eta_{s}\right)  ^{\sum_{\sigma}\hat
{n}_{is,\sigma}} \label{2-0}%
\end{equation}
where $i$ is site index, $s$ denotes each local correlated orbital and
$\eta_{s}$ is the weight of that state. Now the Gutzwiller trial wavefunction
to be used in defining the renormalization factors is
\begin{equation}
\left\vert \tilde{\Psi}\right\rangle _{G}=\left(  \prod_{i}\hat{W}_{i}\hat
{P}_{i,G}\right)  \left\vert \Psi_{0}\right\rangle \label{2-0a}%
\end{equation}
Note, this wavefunction is not supposed to be used to evaluate physical
expectation values of any operator. They are evaluated with Eq. \ref{1-0}
directly, and are thus unique given a Gutzwiller parameter $g$. Of course, the
choice of the weighting operator, $\hat{W}_{i},$ is very flexible, depending
on what kind of physical properties one decides to preserve. In case of a
Hydrogen dimer described by a large basis set of 6-311G and to be studied
here, we use Eq. \ref{2-0} to ensure correct total number of electrons in the
dimer system. The correlated orbital is chosen to be $1s$ while other orbitals
are treated uncorrelated. With these enhanced definitions, the whole averaging
formalism can be written out and $\eta_{s}$ can be determined analytically as
a function of $g$. Details of the calculation are given in Appendix B. Rather
unexpectedly, such a slightly modified averaging scheme gives an exact
description on the multiband Hydrogen dimer system within a Gutzwiller trial
wavefunction. Besides the nice agreement, one might appreciate a piece of
physics embedded in the current approximation. We found that the fugacity
parameter has two solutions. One solution gives orbital occupations identical
to those of the underlying noninteracting wavefunction, a scenario adopted in
the multiband GA formalism, and the other gives the correct orbital occupation
in the Gutzwiller trial wavefunction. The coincidence of the solutions to the
assumption underlying the multiband GA formalism might hint that the weighted
version of the new approximation is intimately related with the multiband GA
formalism that Bunemann et. al derived, just like AA to GA in the single band
case. This fact might also speak loud that the new approximation introduced in
this work could perform better than GA in capturing the strong correlation
physics inherent in the Gutzwiller wavefunction formalism.

\section{Conclusion}

In sum, we have introduced a very simple yet effective approximation towards
the Gutzwiller wavefunction formalism. The simplicity is easily seen by noting
that the renormalization factor of an operator is obtained through a direct
average over nonvanishing coefficients of a Gutzwiller wavefunction
constructed with a density-density type Gutzwiller projector. The
effectiveness is supported by the agreement between the new approximation and
GA, a well-studied approximation towards the Gutzwiller wavefunction
formalism, in the infinite lattice limit of single band systems. Thus the
current approximation provides a new perspective towards GA and its underlying
assumptions. The proof showing their mutual relationship in Appendix A clearly
reveals the grand canonical ensemble nature of GA, which readily prevents it
from being applied to finite systems. This very nature of GA also shows itself
in the counting argument interpretation of GA\cite{Bunemann_GA_EurPhysJB.4.29}
but is not clear from the diagram based
formulation\cite{IDL-MBGA_JPhysCondensMatter.9.7343}. In this work, several
numerical instances are provided to compare the performance of the new
approximation and GA, and to show the asymptotic agreement between these two schemes.

Although it shows an improved performance than GA in single band systems, the
naive averaging, however, does not capture the double occupancy and other
subtle physical quantities well enough. It also leads to the failure to reach
the correct local orbital occupation in a Hydrogen dimer described with a
large basis set, the simplest multiband system to be studied. All these can be
improved however, at least partially, by introducing more fugacity parameters
and physical constraints into the scheme. For example, a simple modification
of introducing a weighting operator like Eq. \ref{2-0} nicely fixes the
multi-orbital Hydrogen dimer problem. Actually, such a simple problem
indicates a possible close relation between the weighted averaging scheme and
the multiband GA formalism. With a more carefully designed weighting factor
and more physical constraints included in the formalism, the performance of
the current approximation could be systematically improved to reach that of
the Gutzwiller wavefunction formalism, with which the current approximation is
built to match.

This new idea is applied most conveniently towards Gutzwiller trial
wavefunction with its Gutzwiller projector commuting with number operators.
Such a density-density type Gutzwiller projector (d-GPJ) might sound inferior
than a more generic Gutzwiller projector (g-GPJ) defined with a general
interaction operator to describe local correlation
effects\cite{MBGA_GONSITE_PhysRevB.57.6896}. This might cast doubt on the
ultimate usefulness of the current scheme to describe real systems. We have to
admit that more variational degrees of freedom introduced in g-GPJ are able to
give ground state energy. But we also believe that d-GPJ might perform with
close quality as g-GPJ. It might also have an additional advantage of being
more convenient in its implementation in practice. To fill us with some
confidence first, we might find a successful application of d-GPJ in capturing
the correct physics. Actually, such a Gutzwiller trial wavefunction has been
applied to the Hubbard model and leads to the well-known t-J Hamiltonian in
the strong correlation limit\cite{HUBBARD_t-J.1990}. The t-J Hamiltonian, as
well as the Heisenberg Hamiltonian as its special case, is the main working
horse towards strong correlation limit and spin dynamics. Second, we believe
that g-GPJ can only be treated in the multiband GA level in practice. Thus a
much fairer comparison would be between the multiband GA and the current
scheme, to compare which one is able to perform better. Then, the answer is
clear that no party wins over the other for sure. One motivation to develop
the formalism involving g-GPJ might be because GA from d-GPJ has to assume
$\left\langle \hat{n}_{\left(  \alpha,\sigma\right)  }\right\rangle
=n_{\left(  \alpha,\sigma\right)  }^{0}$, a serious limitation in applying GA
to multiband systems\cite{IDL-MBGA_JPhysCondensMatter.9.7343}. While GA from
g-GPJ has this constraint removed, it still adopts constraints of similar
nature in its derivation in order to carry out a Feynmann diagrammatic
expansion\cite{MBGA_GONSITE_PhysRevB.57.6896}. Thus, how far multiband GA can
be free from this constraint is unknown to us. In the new approximation
introduced in this work, however, local orbital renormalization inherently
exists which necessarily gives a local orbital occupation differring from its
noninteracting value. One might systematically introduce more and more
physical constraints into the formalism to help guide its outcome to be more
and more physical. One more advantage of the current scheme against GA is that
it is readily applied to Gutzwiller projectors with nonlocal density-density
correlations. This fact is quite easy to notice although its practical
implementation might be an issue.

\begin{acknowledgments}
We are grateful to J. Bunemann and J. Schmalian for helpful discussions and
valuable suggestions. This work is supported by the U.S. Department of Energy,
Office of Basic Energy Sciences, Division of Materials Sciences and
Engineering. Ames Laboratory is operated for the U.S. Department of Energy by
Iowa State University under Contract No. DE-AC02-07CH11358.
\end{acknowledgments}

%

\begin{subappendices}%

\section{APPENDIX A: FROM THE NEW APPROXIMATION TO THE GUTZWILLER
APPROXIMATION ON A SINGLE BAND SYSTEM}

To establish a connection between these two approximations, let's consider the
renormalization coefficient for $\hat{O}\left(  i,j\right)  =c_{i,\sigma
}^{\dagger}c_{j,\sigma}$ with site indices $i\neq j$. Here $\sigma$ denotes a
spin-orbital composite state. Following the spirit leading to Eq. \ref{1-9} to
Eq. \ref{1-9b}, one can write down $\mathcal{F}$ and $\mathcal{B}$ for
$c_{i,\sigma}^{\dagger}c_{j,\sigma}$ as%
\begin{align}
\mathcal{F}  &  \mathcal{=}\frac{\sum_{\left\{  \mathbf{\Gamma}\right\}  }%
\sum_{\left\{  \mathbf{\Gamma}^{\prime}\right\}  }\left(  \prod_{k\neq
i,j}g_{\Gamma_{k}}^{2}\right)  \left(  g_{\Gamma_{i}}g_{\Gamma_{j}}%
g_{\Gamma_{i}^{\prime}}g_{\Gamma_{j}^{\prime}}\right)  \mathcal{S}\left(
\hat{O};\mathbf{\Gamma},\mathbf{\Gamma}^{\prime}\right)  \delta
_{\mathbf{\Gamma}}^{\left(  3\right)  }\delta_{\mathbf{\Gamma}^{\prime}%
}^{\left(  3\right)  }}{\sum_{\left\{  \mathbf{\Gamma}\right\}  }%
\sum_{\left\{  \mathbf{\Gamma}^{\prime}\right\}  }\mathcal{S}\left(  \hat
{O};\mathbf{\Gamma},\mathbf{\Gamma}^{\prime}\right)  \delta_{\mathbf{\Gamma}%
}^{\left(  3\right)  }\delta_{\mathbf{\Gamma}^{\prime}}^{\left(  3\right)  }%
}\\
\mathcal{B}  &  \mathcal{=}\frac{\sum_{\left\{  \mathbf{\Gamma}\right\}
}\left(  \prod_{k}g_{\Gamma_{k}}^{2}\right)  \delta_{\mathbf{\Gamma}}^{\left(
3\right)  }}{\sum_{\left\{  \mathbf{\Gamma}\right\}  }\delta_{\mathbf{\Gamma}%
}^{\left(  3\right)  }}%
\end{align}
Here $\left\{  \mathbf{\Gamma}\right\}  $ and $\left\{  \mathbf{\Gamma
}^{\prime}\right\}  $ denote a complete set of occupation configuration on the
whole lattice, $g_{\Gamma_{i}}$ denotes the $g$ factor at site $i$ with local
Fock state $\Gamma_{i}$. In order to reach GA, let's first relax the $\delta$
constraints, and note the following relations hold%
\begin{align}
\sum_{\left\{  \mathbf{\Gamma}\right\}  }\left(  \prod_{k}g_{\Gamma_{k}}%
^{2}\right)   &  =\left(  \sum_{\Gamma}g_{\Gamma}^{2}\right)  ^{N}\\
\sum_{\left\{  \mathbf{\Gamma}\right\}  }1  &  =\mathcal{D}^{N}%
\end{align}
with $N$ total number of lattice sites and $\mathcal{D}$ dimensionality of the
local Fock space. We thus have
\begin{equation}
\eta=\frac{\mathcal{F}}{\mathcal{B}}=\frac{4}{\left(  \sum_{\Gamma}g_{\Gamma
}^{2}\right)  ^{2}}\left(  \sum_{\Gamma_{i},\Gamma_{j},\Gamma_{i}^{\prime
},\Gamma_{j}^{\prime}}g_{\Gamma_{i}}g_{\Gamma_{j}}g_{\Gamma_{i}^{\prime}%
}g_{\Gamma_{j}^{\prime}}\left\vert \left\langle \Gamma_{i}\right\vert
c_{i,\sigma}^{\mathbf{\dagger}}\left\vert \Gamma_{i}^{\prime}\right\rangle
\right\vert \left\vert \left\langle \Gamma_{j}\right\vert c_{j,\sigma
}\left\vert \Gamma_{j}^{\prime}\right\rangle \right\vert \right)  \label{1-13}%
\end{equation}
This expression for $\eta$ is only intermediate and not fully consistent, as
one can see that $\eta$ fails to vanish as variational parameters $g$ in the
standard Gutzwiller projector approaches 0.

Now let us introduce another approximation that the probability of finding a
given Fock space occupation is a product of probabilities of finding the
specific occupation configuration on each lattice site, or namely,%
\begin{align}
\left\vert \left\langle \left\{  \Gamma\right\}  |\Psi\right\rangle
\right\vert ^{2}  &  =\prod_{k}p_{\Gamma_{k}}\\
\left\vert \left\langle \left\{  \Gamma_{k}\right\}  |\Psi_{0}\right\rangle
\right\vert ^{2}  &  =\prod_{k}p_{\Gamma_{k}}^{0}%
\end{align}
where $p_{\Gamma_{k}},p_{\Gamma_{k}}^{0}$ denote the probability to find a
Fock state $\Gamma_{k}$ on site $k$ for the trial and noninteracting
wavefunction respectively. This is too big a step forward as $\left\vert
\left\langle \left\{  \Gamma\right\}  |\Psi\right\rangle \right\vert ^{2}$ is
normally not decomposeable on lattice sites. This approximation can be
validated only in the infinite spatial dimension limit\cite{comment2}, which
also underlies the Gutzwiller approximation. With this approximation, it is
reasonable to assume
\begin{equation}
g_{\Gamma_{k}}=\sqrt{\frac{p_{\Gamma_{k}}}{p_{\Gamma_{k}}^{0}}} \label{1-14}%
\end{equation}
for the Gutzwiller trial wavefunction defined in Eq. \ref{1-0} by noting that
the local projection operator is site-wise. Feed Eq. \ref{1-14} into Eq.
\ref{1-13} and we get
\begin{equation}
\eta=\bar{z}_{i}\bar{z}_{j}%
\end{equation}
with
\begin{equation}
\bar{z}_{i}=\frac{2}{\left(  \sum_{\Gamma}p_{\Gamma}/p_{\Gamma}^{0}\right)
}\sum_{\Gamma_{i},\Gamma_{i}^{\prime}}\sqrt{\frac{p_{\Gamma_{i}}p_{\Gamma
_{i}^{\prime}}}{p_{\Gamma_{i}}^{0}p_{\Gamma_{i}^{\prime}}^{0}}}\left\vert
\left\langle \Gamma_{i}\right\vert c_{i,\sigma}\left\vert \Gamma_{i}^{\prime
}\right\rangle \right\vert \label{1-15}%
\end{equation}

A mean-field type approximation is now introduced for terms involving
$p_{\Gamma}^{0}$ by replacing those terms with their averages. Again, in the
infinite spatial dimension limit and under some convenience conditions,
$p_{\Gamma}^{0}$ can be expressed in terms of local occupancy, $n_{\sigma}%
^{0}$ on each local state $\sigma$, as
\begin{equation}
p_{\Gamma}^{0}=\prod_{\sigma\in\Gamma}n_{\sigma}^{0}\prod_{\sigma\notin\Gamma
}\left(  1-n_{\sigma}^{0}\right)
\end{equation}
Then there are two averages needed to be calculated,%
\begin{align}
\overline{\sqrt{p^{0}p^{0}}}  &  =\frac{1}{\left(  \frac{\mathcal{D}}%
{2}\right)  }\sum_{\Gamma_{i},\Gamma_{i}^{\prime}}\sqrt{p_{\Gamma_{i}}%
^{0}p_{\Gamma_{i}^{\prime}}^{0}}\left\vert \left\langle \Gamma_{i}\right\vert
c_{i,\sigma}\left\vert \Gamma_{i}^{\prime}\right\rangle \right\vert \\
&  =\frac{1}{\left(  \frac{\mathcal{D}}{2}\right)  }\sum_{\sigma\notin
\Gamma_{i}^{\prime}}\sqrt{p_{\Gamma_{i}}^{0}p_{\Gamma_{i}\cup\sigma}^{0}}\\
&  =\frac{2}{\mathcal{D}}\sqrt{n_{\sigma}^{0}\left(  1-n_{\sigma}^{0}\right)
} \label{1-16a}%
\end{align}
and
\begin{equation}
\overline{p^{0}}=\frac{1}{\mathcal{D}}\sum_{\Gamma}p_{\Gamma}^{0}=\frac
{1}{\mathcal{D}} \label{1-16b}%
\end{equation}
Here $\mathcal{D}$ is the dimensionality of the local Fock space. The
prefactor $2$ in Eq. \ref{1-16a} accounts for the fact that $\Gamma_{i}$ must
not contain state $\sigma$ in it. Plug Eq. \ref{1-16a} and Eq. \ref{1-16b}
back to Eq. \ref{1-15} and one recovers the standard GA definition of $z_{i}$
given in Eq. \ref{1-1a}.

\section{APPENDIX B: THE NEW APPROXIMATION WITH WEIGHTED AVERAGE APPLIED ON
THE MULTIBAND H DIMER SYSTEM}

To make things simple, let's consider a local basis set composed of one
correlated orbital, denoted as $s,$ and $\mathfrak{N}$ uncorrelated orbitals
to describe the H-dimer, all defined as Wannier functions such that they are
orthogonal to each other within and between sites. A molecular orbital is thus
generically created via operator
\begin{equation}
a_{\sigma}^{\dagger}=\sum_{i}h_{is}c_{is,\sigma}^{\dagger}+\sum_{i,\alpha\neq
s}h_{i\alpha}c_{i\alpha,\sigma}^{\dagger}%
\end{equation}
where $c_{i\alpha,\sigma}^{\dagger}$ creates an electron at site $i$ and
orbital $\alpha$ with spin $\sigma,$ or, $c_{i\alpha,\sigma}^{\dagger
}\left\vert 0\right\rangle =\left\vert i\alpha,\sigma\right\rangle $. The
coefficients satisfy the normalization condition%
\begin{equation}
\left\vert h_{1s}\right\vert ^{2}+\sum_{\alpha\neq s}\left\vert h_{1\alpha
}\right\vert ^{2}=\frac{1}{2} \label{2-1}%
\end{equation}
with translational invariance implicitly assumed. That the cross terms
contributing to the normalization vanish comes from the fact that each atomic
orbital is a Wannier function, as mentioned at the beginning of this appendix.
Eq. \ref{2-1} also expresses the electron conservation condition, ensuring
each site has half an electron with a specific spin. The noninteracting
Hartree-Fock wavefunction can be expressed as
\begin{equation}
\left\vert \Psi_{0}\right\rangle =a_{\uparrow}^{\dagger}a_{\downarrow
}^{\dagger}\left\vert 0\right\rangle
\end{equation}
for the H dimer, and the Gutzwiller wavefunction is defined as
\begin{align}
\left\vert \Psi\right\rangle  &  =g^{\sum_{i}\hat{n}_{is\uparrow}\hat
{n}_{is\downarrow}}\left\vert \Psi_{0}\right\rangle \nonumber\\
&  =\left(  g-1\right)  \left\vert h_{1s}\right\vert ^{2}\left\vert
1s\uparrow,1s\downarrow\right\rangle +\left(  g-1\right)  \left\vert
h_{2s}\right\vert ^{2}\left\vert 2s\uparrow,2s\downarrow\right\rangle
+\left\vert \Psi_{0}\right\rangle
\end{align}
with $g$ the Gutzwiller parameter. Expectation value w.r.t the Gutzwiller
wavefunction for any operator can be straightforwardly evaluated.
Specifically, for particle occupations, there are
\begin{align}
\left\langle n_{1s,\sigma}\right\rangle  &  =\frac{\left(  g^{2}-1\right)
\left\vert h_{1s}\right\vert ^{2}+1}{2\left(  g^{2}-1\right)  \left\vert
h_{1s}\right\vert ^{4}+1}\left\vert h_{1s}\right\vert ^{2}\\
\left\langle n_{1\alpha,\sigma}\right\rangle  &  =\frac{1}{2\left(
g^{2}-1\right)  \left\vert h_{1s}\right\vert ^{4}+1}\left\vert h_{1\alpha
}\right\vert ^{2}\text{ for }\alpha\neq s
\end{align}
Obviously, it is easy to test that they satisfy the electron conservation
condition, Eq. \ref{2-1}, with help of that constraint.

For the current approximation with weighted average enhancement, the local
weighting operator is chosen as
\[
\hat{W}_{i}=\eta^{\sum_{\sigma}\hat{n}_{is,\sigma}}%
\]
The expectation value for an operator is considered from its denominator,
wavefunction normalization, and its numerator respectively. For the
normalization factor $\left\langle \Psi|\Psi\right\rangle ,$ there is
\begin{align}
\left\langle \Psi|\Psi\right\rangle  &  =\left\langle \Psi_{0}\right\vert
\eta^{2\sum_{\sigma=\uparrow,\downarrow}\hat{n}_{is,\sigma}}g^{2\sum_{i}%
\hat{n}_{is\uparrow}\hat{n}_{is\downarrow}}\left\vert \Psi_{0}\right\rangle
\nonumber\\
&  \simeq\frac{2\eta^{4}g^{2}+2\eta^{4}+4C_{2\mathfrak{N}}^{1}\eta^{2}+\left(
2C_{2\mathfrak{N}}^{2}+2\mathfrak{N}\right)  }{2\eta^{4}+2\eta^{4}%
+4C_{2\mathfrak{N}}^{1}\eta^{2}+\left(  2C_{2\mathfrak{N}}^{2}+2\mathfrak{N}%
\right)  }\left\langle \Psi_{0}|\Psi_{0}\right\rangle \label{2-2}%
\end{align}
with $C_{n}^{m}$ the usual combinatorial number choosing $m$ elements out of
$n$ elements. In the numerator of the renormalization factor in Eq. \ref{2-2},
each term has clear physical interpretation. $\eta^{4}g^{2}$ corresponds to
two electrons occupying $s$ orbitals on the same site, $\eta^{4}$ has the two
electrons take $s$ orbitals on different sites, $\eta^{2}$ is related to Fock
states with only one electron in $s$ orbitals, while the last term corresponds
to Fock states with no $s$ orbital. The terms in the denominator of the above
prefactor are obtained by ignoring the variational parameter $g.$ What is left
acts as the weight to each term in the numerator, a necessary step to
normalize a weighted average. Similarly, one can write down the numerator of
the expectation value of an operator. For electron occupations, they are
\begin{align}
\left\langle \Psi\right\vert \hat{n}_{1s,\uparrow}\left\vert \Psi
\right\rangle  &  \simeq\frac{\eta^{4}g^{2}+\eta^{4}+C_{2\mathfrak{N}}^{1}%
\eta^{2}}{2\eta^{4}+C_{2\mathfrak{N}}^{1}\eta^{2}}\left\vert h_{1s}\right\vert
^{2}\label{2-2a}\\
\left\langle \Psi\right\vert \hat{n}_{1\alpha,\uparrow}\left\vert
\Psi\right\rangle  &  \simeq\left\vert h_{1\alpha}\right\vert ^{2}\text{ for
}\alpha\neq s \label{2-2b}%
\end{align}
The constraint of conserved electron occupation on a H dimer requires
\begin{equation}
\left\langle \Psi\right\vert \hat{n}_{1s,\uparrow}\left\vert \Psi\right\rangle
+\sum_{\alpha\neq s}\left\langle \Psi\right\vert \hat{n}_{1\alpha,\uparrow
}\left\vert \Psi\right\rangle =\frac{1}{2}\left\langle \Psi|\Psi\right\rangle
\label{2-3}%
\end{equation}
Feed Eq. \ref{2-2}, Eq. \ref{2-2a} and Eq. \ref{2-2b} into Eq. \ref{2-3} and
solve for $\eta,$ and one ends up with two solutions
\begin{align}
\eta^{2}  &  =0\\
\eta^{2}  &  =\frac{\mathfrak{N}\left\vert h_{1s}\right\vert ^{2}}%
{\sum_{\alpha\neq s}\left\vert h_{1\alpha}\right\vert ^{2}} \label{2-4}%
\end{align}

Interestingly, both these two solutions have clear physical interpretations.
$\eta=0$ corresponds to the case where the correlated system has the same
local orbital occupation as the underlying noninteracting wavefunction, which
is what Bunemann's multiband Gutzwiller approximation starts with. The
nontrivial solution of Eq. \ref{2-4} to $\eta$ gives the correct charge
occupation as the rigorous Gutzwiller wavefunction. One can readily verify
this fact by inserting the solution Eq. \ref{2-4} back to the expressions for
$\left\langle n_{1s,\uparrow}\right\rangle $ and $\left\langle n_{1\alpha
,\uparrow}\right\rangle $ and take the constraint of Eq. \ref{2-1}. Actually,
one can further verify that this nontrivial solution renders correct
expressions for any one and two body operators of the H dimer system.%

\end{subappendices}%

\bibliographystyle{apsrev4-1}
\bibliography{introduction_AA}

\begin{thebibliography}{10}%
\makeatletter
\providecommand \@ifxundefined [1]{%
 \ifx #1\undefined \expandafter \@firstoftwo
 \else \expandafter \@secondoftwo
\fi
}%
\providecommand \@ifnum [1]{%
 \ifnum #1\expandafter \@firstoftwo
 \else \expandafter \@secondoftwo
\fi
}%
\providecommand \enquote [1]{``#1''}%
\providecommand \bibnamefont  [1]{#1}%
\providecommand \bibfnamefont [1]{#1}%
\providecommand \citenamefont [1]{#1}%
\providecommand\href[0]{\@sanitize\@href}%
\providecommand\@href[1]{\endgroup\@@startlink{#1}\endgroup\@@href}%
\providecommand\@@href[1]{#1\@@endlink}%
\providecommand \@sanitize [0]{\begingroup\catcode`\&12\catcode`\#12\relax}%
\@ifxundefined \pdfoutput {\@firstoftwo}{%
 \@ifnum{\z@=\pdfoutput}{\@firstoftwo}{\@secondoftwo}%
}{%
 \providecommand\@@startlink[1]{\leavevmode\special{html:<a href="#1">}}%
 \providecommand\@@endlink[0]{\special{html:</a>}}%
}{%
 \providecommand\@@startlink[1]{%
  \leavevmode
  \pdfstartlink
   attr{/Border[0 0 1 ]/H/I/C[0 1 1]}%
   user{/Subtype/Link/A<</Type/Action/S/URI/URI(#1)>>}%
  \relax
 }%
 \providecommand\@@endlink[0]{\pdfendlink}%
}%
\providecommand \url  [0]{\begingroup\@sanitize \@url }%
\providecommand \@url [1]{\endgroup\@href {#1}{\urlprefix}}%
\providecommand \urlprefix [0]{URL }%
\providecommand \Eprint[0]{\href }%
\@ifxundefined \urlstyle {%
  \providecommand \doi [1]{doi:\discretionary{}{}{}#1}%
}{%
  \providecommand \doi [0]{doi:\discretionary{}{}{}\begingroup
  \urlstyle{rm}\Url }%
}%
\providecommand \doibase [0]{http://dx.doi.org/}%
\providecommand \Doi[1]{\href{\doibase#1}}%
\providecommand \bibAnnote [3]{%
  \BibitemShut{#1}%
  \begin{quotation}\noindent
    \textsc{Key:}\ #2\\\textsc{Annotation:}\ #3%
  \end{quotation}%
}%
\providecommand \bibAnnoteFile [2]{%
  \IfFileExists{#2}{\bibAnnote {#1} {#2} {\input{#2}}}{}%
}%
\providecommand \typeout [0]{\immediate \write \m@ne }%
\providecommand \selectlanguage [0]{\@gobble}%
\providecommand \bibinfo [0]{\@secondoftwo}%
\providecommand \bibfield [0]{\@secondoftwo}%
\providecommand \translation [1]{[#1]}%
\providecommand \BibitemOpen[0]{}%
\providecommand \bibitemStop [0]{}%
\providecommand \bibitemNoStop [0]{.\EOS\space}%
\providecommand \EOS [0]{\spacefactor3000\relax}%
\providecommand \BibitemShut [1]{\csname bibitem#1\endcsname}%
\bibitem{CORRELEC.2003}%
  \BibitemOpen
  \bibfield{author}{%
  \bibinfo {author} {\bibfnamefont{P.}~\bibnamefont{Fazekas}},\ }%
  \emph{\bibinfo {title} {Lecture Notes on Electron Corrrelation and
  Magnetism}},\ \bibinfo {edition} {2nd}\ ed.\ (\bibinfo {publisher} {World
  Scientific},\ \bibinfo {address} {Singapore},\ \bibinfo {year} {2003})%
  \bibAnnoteFile{NoStop}{CORRELEC.2003}%
\bibitem{HighTC.1997}%
  \BibitemOpen
  \bibfield{author}{%
  \bibinfo {author} {\bibfnamefont{P.~W.}\ \bibnamefont{Anderson}},\ }%
  \emph{\bibinfo {title} {The Theory of Superconductivity in the High-Tc
  Cuprate Superconductors}},\ \bibinfo {edition} {1st}\ ed.\ (\bibinfo
  {publisher} {Princeton University Press},\ \bibinfo {address} {New Jersey,
  USA},\ \bibinfo {year} {1997})%
  \bibAnnoteFile{NoStop}{HighTC.1997}%
\bibitem{HeavyFermion.2007}%
  \BibitemOpen
  \bibfield{author}{%
  \bibinfo {author} {\bibfnamefont{P.}~\bibnamefont{Coleman}},\ }%
  \emph{\bibinfo {title} {Heavy Fermions: electrons at the edge of magnetism,
  in Handbook of Magnetism and Advanced Magnetic Materials, Vol 1: Fundamentals
  and Theory, ed. by H. Kronmuller and S. Parkin}},\ \bibinfo {edition} {1st}\
  ed.\ (\bibinfo {publisher} {J. Wiley and Sons},\ \bibinfo {address} {New
  Jersey, USA},\ \bibinfo {year} {2007})\ pp.\ \bibinfo {pages} {95--148}%
  \bibAnnoteFile{NoStop}{HeavyFermion.2007}%
\bibitem{OPTIC_RevModPhys.83.471}%
  \BibitemOpen
  \bibfield{author}{%
  \bibinfo {author} {\bibfnamefont{D.~N.}\ \bibnamefont{Basov}}, \bibinfo
  {author} {\bibfnamefont{R.~D.}\ \bibnamefont{Averitt}}, \bibinfo {author}
  {\bibfnamefont{D.}~\bibnamefont{van~der Marel}}, \bibinfo {author}
  {\bibfnamefont{M.}~\bibnamefont{Dressel}},\ and\ \bibinfo {author}
  {\bibfnamefont{K.}~\bibnamefont{Haule}},\ }%
  \bibfield{journal}{%
  \bibinfo {journal} {Rev. Mod. Phys.}\ }%
  \textbf{\bibinfo {volume} {83}},\ \bibinfo {pages} {471} (\bibinfo {year}
  {2011})%
  \bibAnnoteFile{NoStop}{OPTIC_RevModPhys.83.471}%
\bibitem{PhysRevLett.92.226402}%
  \BibitemOpen
  \bibfield{author}{%
  \bibinfo {author} {\bibfnamefont{O.}~\bibnamefont{Parcollet}}, \bibinfo
  {author} {\bibfnamefont{G.}~\bibnamefont{Biroli}},\ and\ \bibinfo {author}
  {\bibfnamefont{G.}~\bibnamefont{Kotliar}},\ }%
  \bibfield{journal}{%
  \bibinfo {journal} {Phys. Rev. Lett.}\ }%
  \textbf{\bibinfo {volume} {92}},\ \bibinfo {pages} {226402} (\bibinfo {year}
  {2004})%
  \bibAnnoteFile{NoStop}{PhysRevLett.92.226402}%
\bibitem{PhysRevLett.94.127003}%
  \BibitemOpen
  \bibfield{author}{%
  \bibinfo {author} {\bibfnamefont{J.}~\bibnamefont{Liu}}, \bibinfo {author}
  {\bibfnamefont{J.}~\bibnamefont{Schmalian}},\ and\ \bibinfo {author}
  {\bibfnamefont{N.}~\bibnamefont{Trivedi}},\ }%
  \bibfield{journal}{%
  \bibinfo {journal} {Phys. Rev. Lett.}\ }%
  \textbf{\bibinfo {volume} {94}},\ \bibinfo {pages} {127003} (\bibinfo {year}
  {2005})%
  \bibAnnoteFile{NoStop}{PhysRevLett.94.127003}%
\bibitem{QUANTHEORY.2003}%
  \BibitemOpen
  \bibfield{author}{%
  \bibinfo {author} {\bibfnamefont{A.~L.}\ \bibnamefont{Fetter}}\ and\ \bibinfo
  {author} {\bibfnamefont{J.~D.}\ \bibnamefont{Walecka}},\ }%
  \emph{\bibinfo {title} {Quantum Theory of Many Particle Systems}},\ \bibinfo
  {edition} {1st}\ ed.\ (\bibinfo {publisher} {Dover Publications},\ \bibinfo
  {address} {New York, US},\ \bibinfo {year} {2003})%
  \bibAnnoteFile{NoStop}{QUANTHEORY.2003}%
\bibitem{SpinFluc.2008}%
  \BibitemOpen
  \bibfield{author}{%
  \bibinfo {author} {\bibfnamefont{D.~P.}\ \bibnamefont{A.~Chubukov}}\ and\
  \bibinfo {author} {\bibfnamefont{J.}~\bibnamefont{Schmalian}},\ }%
  \emph{\bibinfo {title} {A Spin Fluctuation Model for d-wave Superconductors,
  in The Physics of Superconductors, ed. K.H. Benneman and J. B. Ketterson}},\
  \bibinfo {edition} {2nd}\ ed.\ (\bibinfo {publisher} {Springer},\ \bibinfo
  {address} {Berlin, Germany},\ \bibinfo {year} {2008})\ pp.\ \bibinfo {pages}
  {1349--1407}%
  \bibAnnoteFile{NoStop}{SpinFluc.2008}%
\bibitem{AnnPhys.193.206}%
  \BibitemOpen
  \bibfield{author}{%
  \bibinfo {author} {\bibfnamefont{N.~E.}\ \bibnamefont{Bickers}}\ and\
  \bibinfo {author} {\bibfnamefont{D.~J.}\ \bibnamefont{Scalapino}},\ }%
  \bibfield{journal}{%
  \bibinfo {journal} {Ann. Phys.}\ }%
  \textbf{\bibinfo {volume} {193}},\ \bibinfo {pages} {206} (\bibinfo {year}
  {1989})%
  \bibAnnoteFile{NoStop}{AnnPhys.193.206}%
\bibitem{SolidStateCommun.82.311}%
  \BibitemOpen
  \bibfield{author}{%
  \bibinfo {author} {\bibfnamefont{C.~X.}\ \bibnamefont{Chen}}\ and\ \bibinfo
  {author} {\bibfnamefont{N.~E.}\ \bibnamefont{Bickers}},\ }%
  \bibfield{journal}{%
  \bibinfo {journal} {Solid State Commun.}\ }%
  \textbf{\bibinfo {volume} {82}},\ \bibinfo {pages} {311} (\bibinfo {year}
  {1992})%
  \bibAnnoteFile{NoStop}{SolidStateCommun.82.311}%
\bibitem{Parquet.2004}%
  \BibitemOpen
  \bibfield{author}{%
  \bibinfo {author} {\bibfnamefont{N.~E.}\ \bibnamefont{Bickers}},\ }%
  \emph{\bibinfo {title} {in Theoretical Methods for Strongly Correlated
  Electrons, edited by D. Senechal, A. Tremblay, and C. Bour-bonnais}}\
  (\bibinfo {publisher} {Springer-Verlag},\ \bibinfo {address} {New York,
  USA},\ \bibinfo {year} {2004})\ p.\ \bibinfo {pages} {237}%
  \bibAnnoteFile{NoStop}{Parquet.2004}%
\bibitem{ED_PhysRev.135.A640}%
  \BibitemOpen
  \bibfield{author}{%
  \bibinfo {author} {\bibfnamefont{J.~C.}\ \bibnamefont{Bonner}}\ and\ \bibinfo
  {author} {\bibfnamefont{M.~E.}\ \bibnamefont{Fisher}},\ }%
  \bibfield{journal}{%
  \bibinfo {journal} {Phys. Rev.}\ }%
  \textbf{\bibinfo {volume} {135}},\ \bibinfo {pages} {A640} (\bibinfo {year}
  {1964})%
  \bibAnnoteFile{NoStop}{ED_PhysRev.135.A640}%
\bibitem{1st_QMC_PhysRevB.26.5033}%
  \BibitemOpen
  \bibfield{author}{%
  \bibinfo {author} {\bibfnamefont{J.~E.}\ \bibnamefont{Hirsch}}, \bibinfo
  {author} {\bibfnamefont{R.~L.}\ \bibnamefont{Sugar}}, \bibinfo {author}
  {\bibfnamefont{D.~J.}\ \bibnamefont{Scalapino}},\ and\ \bibinfo {author}
  {\bibfnamefont{R.}~\bibnamefont{Blankenbecler}},\ }%
  \bibfield{journal}{%
  \bibinfo {journal} {Phys. Rev. B}\ }%
  \textbf{\bibinfo {volume} {26}},\ \bibinfo {pages} {5033} (\bibinfo {year}
  {1982})%
  \bibAnnoteFile{NoStop}{1st_QMC_PhysRevB.26.5033}%
\bibitem{DMFT_RevModPhys.68.13}%
  \BibitemOpen
  \bibfield{author}{%
  \bibinfo {author} {\bibfnamefont{A.}~\bibnamefont{Georges}}, \bibinfo
  {author} {\bibfnamefont{G.}~\bibnamefont{Kotliar}}, \bibinfo {author}
  {\bibfnamefont{W.}~\bibnamefont{Krauth}},\ and\ \bibinfo {author}
  {\bibfnamefont{M.~J.}\ \bibnamefont{Rozenberg}},\ }%
  \bibfield{journal}{%
  \bibinfo {journal} {Rev. Mod. Phys.}\ }%
  \textbf{\bibinfo {volume} {68}},\ \bibinfo {pages} {13} (\bibinfo {year}
  {1996})%
  \bibAnnoteFile{NoStop}{DMFT_RevModPhys.68.13}%
\bibitem{CDMFT_PhysRevLett.87.186401}%
  \BibitemOpen
  \bibfield{author}{%
  \bibinfo {author} {\bibfnamefont{G.}~\bibnamefont{Kotliar}}, \bibinfo
  {author} {\bibfnamefont{S.~Y.}\ \bibnamefont{Savrasov}}, \bibinfo {author}
  {\bibfnamefont{G.}~\bibnamefont{P\'alsson}},\ and\ \bibinfo {author}
  {\bibfnamefont{G.}~\bibnamefont{Biroli}},\ }%
  \bibfield{journal}{%
  \bibinfo {journal} {Phys. Rev. Lett.}\ }%
  \textbf{\bibinfo {volume} {87}},\ \bibinfo {pages} {186401} (\bibinfo {year}
  {2001})%
  \bibAnnoteFile{NoStop}{CDMFT_PhysRevLett.87.186401}%
\bibitem{DCA_RevModPhys.77.1027}%
  \BibitemOpen
  \bibfield{author}{%
  \bibinfo {author} {\bibfnamefont{T.}~\bibnamefont{Maier}}, \bibinfo {author}
  {\bibfnamefont{M.}~\bibnamefont{Jarrell}}, \bibinfo {author}
  {\bibfnamefont{T.}~\bibnamefont{Pruschke}},\ and\ \bibinfo {author}
  {\bibfnamefont{M.~H.}\ \bibnamefont{Hettler}},\ }%
  \bibfield{journal}{%
  \bibinfo {journal} {Rev. Mod. Phys.}\ }%
  \textbf{\bibinfo {volume} {77}},\ \bibinfo {pages} {1027} (\bibinfo {year}
  {2005})%
  \bibAnnoteFile{NoStop}{DCA_RevModPhys.77.1027}%
\bibitem{NRG_RevModPhys.47.773}%
  \BibitemOpen
  \bibfield{author}{%
  \bibinfo {author} {\bibfnamefont{K.~G.}\ \bibnamefont{Wilson}},\ }%
  \bibfield{journal}{%
  \bibinfo {journal} {Rev. Mod. Phys.}\ }%
  \textbf{\bibinfo {volume} {47}},\ \bibinfo {pages} {773} (\bibinfo {year}
  {1975})%
  \bibAnnoteFile{NoStop}{NRG_RevModPhys.47.773}%
\bibitem{NRG_RevModPhys.80.395}%
  \BibitemOpen
  \bibfield{author}{%
  \bibinfo {author} {\bibfnamefont{R.}~\bibnamefont{Bulla}}, \bibinfo {author}
  {\bibfnamefont{T.~A.}\ \bibnamefont{Costi}},\ and\ \bibinfo {author}
  {\bibfnamefont{T.}~\bibnamefont{Pruschke}},\ }%
  \bibfield{journal}{%
  \bibinfo {journal} {Rev. Mod. Phys.}\ }%
  \textbf{\bibinfo {volume} {80}},\ \bibinfo {pages} {395} (\bibinfo {year}
  {2008})%
  \bibAnnoteFile{NoStop}{NRG_RevModPhys.80.395}%
\bibitem{DMRG_PhysRevLett.69.2863}%
  \BibitemOpen
  \bibfield{author}{%
  \bibinfo {author} {\bibfnamefont{S.~R.}\ \bibnamefont{White}},\ }%
  \bibfield{journal}{%
  \bibinfo {journal} {Phys. Rev. Lett.}\ }%
  \textbf{\bibinfo {volume} {69}},\ \bibinfo {pages} {2863} (\bibinfo {year}
  {1992})%
  \bibAnnoteFile{NoStop}{DMRG_PhysRevLett.69.2863}%
\bibitem{QUANMECH.2000}%
  \BibitemOpen
  \bibfield{author}{%
  \bibinfo {author} {\bibfnamefont{J.}~\bibnamefont{Zeng}},\ }%
  \emph{\bibinfo {title} {Quantum Mechanics, Volume I}},\ \bibinfo {edition}
  {3rd}\ ed.\ (\bibinfo {publisher} {Science Press},\ \bibinfo {address}
  {Beijing, China},\ \bibinfo {year} {2000})%
  \bibAnnoteFile{NoStop}{QUANMECH.2000}%
\bibitem{HK_PhysRev.136.B864}%
  \BibitemOpen
  \bibfield{author}{%
  \bibinfo {author} {\bibfnamefont{P.}~\bibnamefont{Hohenberg}}\ and\ \bibinfo
  {author} {\bibfnamefont{W.}~\bibnamefont{Kohn}},\ }%
  \bibfield{journal}{%
  \Doi{10.1103/PhysRev.136.B864}{\bibinfo {journal} {Phys. Rev.}}\ }%
  \textbf{\bibinfo {volume} {136}},\ \bibinfo {pages} {B864} (\bibinfo {month}
  {Nov}\ \bibinfo {year} {1964}),\
  \url{http://link.aps.org/doi/10.1103/PhysRev.136.B864}%
  \bibAnnoteFile{NoStop}{HK_PhysRev.136.B864}%
\bibitem{KS_PhysRev.140.A1133}%
  \BibitemOpen
  \bibfield{author}{%
  \bibinfo {author} {\bibfnamefont{W.}~\bibnamefont{Kohn}}\ and\ \bibinfo
  {author} {\bibfnamefont{L.~J.}\ \bibnamefont{Sham}},\ }%
  \bibfield{journal}{%
  \Doi{10.1103/PhysRev.140.A1133}{\bibinfo {journal} {Phys. Rev.}}\ }%
  \textbf{\bibinfo {volume} {140}},\ \bibinfo {pages} {A1133} (\bibinfo {month}
  {Nov}\ \bibinfo {year} {1965}),\
  \url{http://link.aps.org/doi/10.1103/PhysRev.140.A1133}%
  \bibAnnoteFile{NoStop}{KS_PhysRev.140.A1133}%
\bibitem{RVB_PhysRevLett.58.2790}%
  \BibitemOpen
  \bibfield{author}{%
  \bibinfo {author} {\bibfnamefont{P.~W.}\ \bibnamefont{Anderson}}, \bibinfo
  {author} {\bibfnamefont{G.}~\bibnamefont{Baskaran}}, \bibinfo {author}
  {\bibfnamefont{Z.}~\bibnamefont{Zou}},\ and\ \bibinfo {author}
  {\bibfnamefont{T.}~\bibnamefont{Hsu}},\ }%
  \bibfield{journal}{%
  \bibinfo {journal} {Phys. Rev. Lett.}\ }%
  \textbf{\bibinfo {volume} {58}},\ \bibinfo {pages} {2790} (\bibinfo {year}
  {1987})%
  \bibAnnoteFile{NoStop}{RVB_PhysRevLett.58.2790}%
\bibitem{RVB_Science.235.1196}%
  \BibitemOpen
  \bibfield{author}{%
  \bibinfo {author} {\bibfnamefont{P.~W.}\ \bibnamefont{Anderson}},\ }%
  \bibfield{journal}{%
  \bibinfo {journal} {Science}\ }%
  \textbf{\bibinfo {volume} {235}},\ \bibinfo {pages} {1196} (\bibinfo {year}
  {1987})%
  \bibAnnoteFile{NoStop}{RVB_Science.235.1196}%
\bibitem{PhysRevLett.10.159}%
  \BibitemOpen
  \bibfield{author}{%
  \bibinfo {author} {\bibfnamefont{M.~C.}\ \bibnamefont{Gutzwiller}},\ }%
  \bibfield{journal}{%
  \bibinfo {journal} {Phys. Rev. Lett.}\ }%
  \textbf{\bibinfo {volume} {10}},\ \bibinfo {pages} {159} (\bibinfo {year}
  {1963})%
  \bibAnnoteFile{NoStop}{PhysRevLett.10.159}%
\bibitem{ProcRoySocA.276.238}%
  \BibitemOpen
  \bibfield{author}{%
  \bibinfo {author} {\bibfnamefont{J.}~\bibnamefont{Hubbard}},\ }%
  \bibfield{journal}{%
  \bibinfo {journal} {Proc. Roy. Soc. A}\ }%
  \textbf{\bibinfo {volume} {276}},\ \bibinfo {pages} {238} (\bibinfo {year}
  {1963})%
  \bibAnnoteFile{NoStop}{ProcRoySocA.276.238}%
\bibitem{PhysRev.134.A923}%
  \BibitemOpen
  \bibfield{author}{%
  \bibinfo {author} {\bibfnamefont{M.~C.}\ \bibnamefont{Gutzwiller}},\ }%
  \bibfield{journal}{%
  \bibinfo {journal} {Phys. Rev.}\ }%
  \textbf{\bibinfo {volume} {134}},\ \bibinfo {pages} {A923} (\bibinfo {year}
  {1964})%
  \bibAnnoteFile{NoStop}{PhysRev.134.A923}%
\bibitem{HUBBARD_t-J_PhysRevB.37.533}%
  \BibitemOpen
  \bibfield{author}{%
  \bibinfo {author} {\bibfnamefont{J.}~\bibnamefont{Spa\l{}ek}},\ }%
  \bibfield{journal}{%
  \bibinfo {journal} {Phys. Rev. B}\ }%
  \textbf{\bibinfo {volume} {37}},\ \bibinfo {pages} {533} (\bibinfo {year}
  {1988})%
  \bibAnnoteFile{NoStop}{HUBBARD_t-J_PhysRevB.37.533}%
\bibitem{HUBBARD_t-J.1990}%
  \BibitemOpen
  \bibfield{author}{%
  \bibinfo {author} {\bibfnamefont{e.~a.}\ \bibnamefont{A.~P.~Balachandran}},\
  }%
  \emph{\bibinfo {title} {Hubbard model and anyon superconductivity}}\
  (\bibinfo {publisher} {World Scientific Publishing Co.},\ \bibinfo {address}
  {Singapore},\ \bibinfo {year} {1990})%
  \bibAnnoteFile{NoStop}{HUBBARD_t-J.1990}%
\bibitem{RevModPhys.56.99}%
  \BibitemOpen
  \bibfield{author}{%
  \bibinfo {author} {\bibfnamefont{D.}~\bibnamefont{Vollhardt}},\ }%
  \bibfield{journal}{%
  \bibinfo {journal} {Rev. Mod. Phys.}\ }%
  \textbf{\bibinfo {volume} {56}},\ \bibinfo {pages} {99} (\bibinfo {year}
  {1984})%
  \bibAnnoteFile{NoStop}{RevModPhys.56.99}%
\bibitem{Ogawa_GA_ProgTheorPhys.53.614}%
  \BibitemOpen
  \bibfield{author}{%
  \bibinfo {author} {\bibfnamefont{K.~K.}\ \bibnamefont{Tohru~Ogawa}}\ and\
  \bibinfo {author} {\bibfnamefont{T.}~\bibnamefont{Matsubara}},\ }%
  \bibfield{journal}{%
  \bibinfo {journal} {Prog. Theor. Phys.}\ }%
  \textbf{\bibinfo {volume} {53}},\ \bibinfo {pages} {614} (\bibinfo {year}
  {1975})%
  \bibAnnoteFile{NoStop}{Ogawa_GA_ProgTheorPhys.53.614}%
\bibitem{Bunemann_GA_EurPhysJB.4.29}%
  \BibitemOpen
  \bibfield{author}{%
  \bibinfo {author} {\bibfnamefont{J.}~\bibnamefont{Bumenann}},\ }%
  \bibfield{journal}{%
  \bibinfo {journal} {Eur. Phys. J. B}\ }%
  \textbf{\bibinfo {volume} {4}},\ \bibinfo {pages} {29} (\bibinfo {year}
  {1998})%
  \bibAnnoteFile{NoStop}{Bunemann_GA_EurPhysJB.4.29}%
\bibitem{IDL-MBGA_JPhysCondensMatter.9.7343}%
  \BibitemOpen
  \bibfield{author}{%
  \bibinfo {author} {\bibfnamefont{J.}~\bibnamefont{Bunemann}}, \bibinfo
  {author} {\bibfnamefont{F.}~\bibnamefont{Gebhard}},\ and\ \bibinfo {author}
  {\bibfnamefont{W.}~\bibnamefont{Weber}},\ }%
  \bibfield{journal}{%
  \bibinfo {journal} {Journal of Physics: Condensed Matter}\ }%
  \textbf{\bibinfo {volume} {9}},\ \bibinfo {pages} {7343} (\bibinfo {year}
  {1997})%
  \bibAnnoteFile{NoStop}{IDL-MBGA_JPhysCondensMatter.9.7343}%
\bibitem{PhysRevLett.62.324}%
  \BibitemOpen
  \bibfield{author}{%
  \bibinfo {author} {\bibfnamefont{W.}~\bibnamefont{Metzner}}\ and\ \bibinfo
  {author} {\bibfnamefont{D.}~\bibnamefont{Vollhardt}},\ }%
  \bibfield{journal}{%
  \bibinfo {journal} {Phys. Rev. Lett.}\ }%
  \textbf{\bibinfo {volume} {62}},\ \bibinfo {pages} {324} (\bibinfo {year}
  {1989})%
  \bibAnnoteFile{NoStop}{PhysRevLett.62.324}%
\bibitem{PhysRevB.37.7382}%
  \BibitemOpen
  \bibfield{author}{%
  \bibinfo {author} {\bibfnamefont{W.}~\bibnamefont{Metzner}}\ and\ \bibinfo
  {author} {\bibfnamefont{D.}~\bibnamefont{Vollhardt}},\ }%
  \bibfield{journal}{%
  \bibinfo {journal} {Phys. Rev. B}\ }%
  \textbf{\bibinfo {volume} {37}},\ \bibinfo {pages} {7382} (\bibinfo {year}
  {1988})%
  \bibAnnoteFile{NoStop}{PhysRevB.37.7382}%
\bibitem{MBGA_GONSITE_PhysRevB.57.6896}%
  \BibitemOpen
  \bibfield{author}{%
  \bibinfo {author} {\bibfnamefont{J.}~\bibnamefont{B\"unemann}}, \bibinfo
  {author} {\bibfnamefont{W.}~\bibnamefont{Weber}},\ and\ \bibinfo {author}
  {\bibfnamefont{F.}~\bibnamefont{Gebhard}},\ }%
  \bibfield{journal}{%
  \bibinfo {journal} {Phys. Rev. B}\ }%
  \textbf{\bibinfo {volume} {57}},\ \bibinfo {pages} {6896} (\bibinfo {year}
  {1998})%
  \bibAnnoteFile{NoStop}{MBGA_GONSITE_PhysRevB.57.6896}%
\bibitem{JPhysSocJapan.65.1056}%
  \BibitemOpen
  \bibfield{author}{%
  \bibinfo {author} {\bibfnamefont{T.}~\bibnamefont{Okabe}},\ }%
  \bibfield{journal}{%
  \bibinfo {journal} {J. Phys. Soc. Japan}\ }%
  \textbf{\bibinfo {volume} {65}},\ \bibinfo {pages} {1056} (\bibinfo {year}
  {1996})%
  \bibAnnoteFile{NoStop}{JPhysSocJapan.65.1056}%
\bibitem{Fabrizio_MBGA_PhysRevB.76.165110}%
  \BibitemOpen
  \bibfield{author}{%
  \bibinfo {author} {\bibfnamefont{M.}~\bibnamefont{Fabrizio}},\ }%
  \bibfield{journal}{%
  \bibinfo {journal} {Phys. Rev. B}\ }%
  \textbf{\bibinfo {volume} {76}},\ \bibinfo {pages} {165110} (\bibinfo {year}
  {2007})%
  \bibAnnoteFile{NoStop}{Fabrizio_MBGA_PhysRevB.76.165110}%
\bibitem{BR-MIT_PhysRevB.2.4302}%
  \BibitemOpen
  \bibfield{author}{%
  \bibinfo {author} {\bibfnamefont{W.~F.}\ \bibnamefont{Brinkman}}\ and\
  \bibinfo {author} {\bibfnamefont{T.~M.}\ \bibnamefont{Rice}},\ }%
  \bibfield{journal}{%
  \bibinfo {journal} {Phys. Rev. B}\ }%
  \textbf{\bibinfo {volume} {2}},\ \bibinfo {pages} {4302} (\bibinfo {year}
  {1970})%
  \bibAnnoteFile{NoStop}{BR-MIT_PhysRevB.2.4302}%
\bibitem{OrbitSelecMT_EurPhyJB.25.191}%
  \BibitemOpen
  \bibfield{author}{%
  \bibinfo {author} {\bibfnamefont{V.}~\bibnamefont{Anisimov}}, \bibinfo
  {author} {\bibfnamefont{I.}~\bibnamefont{Nekrasov}}, \bibinfo {author}
  {\bibfnamefont{D.}~\bibnamefont{Kondakov}}, \bibinfo {author}
  {\bibfnamefont{T.}~\bibnamefont{Rice}},\ and\ \bibinfo {author}
  {\bibfnamefont{M.}~\bibnamefont{Sigrist}},\ }%
  \bibfield{journal}{%
  \Doi{10.1140/epjb/e20020021}{\bibinfo {journal} {The European Physical
  Journal B - Condensed Matter and Complex Systems}}\ }%
  \textbf{\bibinfo {volume} {25}},\ \bibinfo {pages} {191} (\bibinfo {year}
  {2002}),\ ISSN \bibinfo {issn} {1434-6028},\
  \url{http://dx.doi.org/10.1140/epjb/e20020021}%
  \bibAnnoteFile{NoStop}{OrbitSelecMT_EurPhyJB.25.191}%
\bibitem{CCSD.2009}%
  \BibitemOpen
  \bibfield{author}{%
  \bibinfo {author} {\bibfnamefont{I.}~\bibnamefont{Shavitt}}\ and\ \bibinfo
  {author} {\bibfnamefont{R.~J.}\ \bibnamefont{Bartlett}},\ }%
  \emph{\bibinfo {title} {Many-Body Methods in Chemistry and Physics: MBPT and
  Coupled-Cluster Theory}},\ \bibinfo {edition} {1st}\ ed.\ (\bibinfo
  {publisher} {Cambridge University Press},\ \bibinfo {address} {Cambridge,
  UK},\ \bibinfo {year} {2009})%
  \bibAnnoteFile{NoStop}{CCSD.2009}%
\bibitem{GA_GWF_Fill_PhysRevB.37.7382}%
  \BibitemOpen
  \bibfield{author}{%
  \bibinfo {author} {\bibfnamefont{W.}~\bibnamefont{Metzner}}\ and\ \bibinfo
  {author} {\bibfnamefont{D.}~\bibnamefont{Vollhardt}},\ }%
  \bibfield{journal}{%
  \bibinfo {journal} {Phys. Rev. B}\ }%
  \textbf{\bibinfo {volume} {37}},\ \bibinfo {pages} {7382} (\bibinfo {year}
  {1988})%
  \bibAnnoteFile{NoStop}{GA_GWF_Fill_PhysRevB.37.7382}%
\bibitem{comment2}%
  \BibitemOpen
  \bibinfo {note} {This statement can be formally argued in the following way.
  Introduce a bosonic operator
  $\hat{B}_{\Gamma}^{\dagger}=\prod_{\sigma\in\Gamma}\hat{n}_{\sigma}\prod
  _{\sigma\notin\Gamma}\left( 1-\hat{n}_{\sigma}\right) $ for projecting out a
  Fock state $\Gamma$ on each lattice site and the occupation probability for a
  given occupation configuration $\left\{ \Gamma_{i}\right\} $ can be expressed
  as $P=\left\langle \Psi\right\vert \prod_{i} \hat{B}_{i,\Gamma_{i}
  }^{\dagger}\left\vert \Psi\right\rangle $. This expression is then
  decomposeable in site indices in the infinite dimension limit.}%
  \bibAnnoteFile{Stop}{comment2}%
\end{thebibliography}%

\end{document}